\documentclass[12pt,letterpaper,a4paper]{article}
\usepackage[includeheadfoot,
            marginratio={1:1,1:1},
            width=480pt,
            height=755pt,]{geometry}
\usepackage{amsmath}
\usepackage{amsfonts}
\usepackage{amssymb}
\usepackage{graphicx}
\usepackage{empheq}
\usepackage{subfigure}
\usepackage{color}
\usepackage{hyperref}
\usepackage{enumerate}

\usepackage{float}
\restylefloat{table}
\restylefloat{figure}
\newcommand{\nc}{\newcommand}
\nc{\beq}{\begin{equation}}
\nc{\eeq}{\end{equation}}
\nc{\bea}{\begin{eqnarray}}
\nc{\eea}{\end{eqnarray}}
\def\ov{\overline}

\def\f{\frac}
\newdimen\csize\csize=1.5ex
\def\young#1{\tiny\vcenter{\hbox{\vrule\vtop{\hrule
  \offinterlineskip\halign{&\vbox
  {\hbox to\csize {\strut\hss##\hss\vrule}\hrule}\cr#1 \crcr}}}}}

\begin{document}

%\vspace*{-1.5cm}
%\begin{flushright}
%  {\small
%  MPP-2013-15\\
%  }
%\end{flushright}

\vspace{1.5cm}
\begin{center}
{\LARGE
Cosmological observables in multi-field inflation with a non-flat field space}
\vspace{0.4cm}

\end{center}

\vspace{0.35cm}
\begin{center}
 Xin Gao$^{\dag,\ddag}$\footnote{Email: xingao@vt.edu}, Tianjun Li$^{\dag, \flat}$\footnote{Email: tli@itp.ac.cn}  and Pramod Shukla$^\sharp$\footnote{Email: pkshukla@to.infn.it}
\end{center}

\vspace{0.1cm}
\begin{center}
{\it
%$^{1}$ Max-Planck-Institut f\"ur Physik (Werner-Heisenberg-Institut), \\
%   F\"ohringer Ring 6,  80805 M\"unchen, Germany  \\
%\vspace{0.4cm}

$^{\dag}$ State Key Laboratory of Theoretical Physics
and Kavli Institute for Theoretical Physics China (KITPC),
      Institute of Theoretical Physics, Chinese Academy of Sciences,\\
Beijing 100190, P. R. China \\
\vspace{0.4cm}
$^{\ddag}$ Department of Physics, Robeson Hall, 0435,  Virginia Tech,  \\
850 West Campus Drive, Blacksburg, VA 24061, USA\\
\vspace{0.4cm}
$^{\flat}$ School of Physical Electronics,
University of Electronic Science and Technology of China,
Chengdu 610054, P. R. China \\
\vspace{0.4cm}
%$^2$ Consortium for Fundamental Physics, Physics Department, Lancaster University, LA1
%4YB, United Kingdom\\
%\vspace{0.4cm}
$^{\sharp}$ Universit\'a di Torino, Dipartimento di Fisica and I.N.F.N. - sezione di Torino \\
Via P. Giuria 1, I-10125 Torino, Italy
}

\vspace{0.2cm}

\vspace{0.5cm}
\end{center}

\vspace{1cm}

%%%%%%%%%%%%%%%%%%%%%%%%%%%%%%%%%%%%%%%%%%%%%%%
%%%%%%%%%%%%%%%%%%%%%%%%%%%%%%%%%%%%%%%%%%%%%%%
%%%%%%%%%%%%%%%%%%%%%%%%%%%%%%%%%%%%%%%%%%%%%%%
%%%%%%%%%%%%%%%%%%%%%%%%%%%%%%%%%%%%%%%%%%%%%%%
%%%%%%%%%%%%%%%%%%%%%%%%%%%%%%%%%%%%%%%%%%%%%%%
%%%%%%%%%%%%%%%%%%%%%%%%%%%%%%%%%%%%%%%%%%%%%%%
%%%%%%%%%%%%%%%%%%%%%%%%%%%%%%%%%%%%%%%%%%%%%%%
%%%%%%%%%%%%%%%%%%%%%%%%%%%%%%%%%%%%%%%%%%%%%%%

\begin{abstract}
Using $\delta N$ formalism, in the context of a generic multi-field inflation driven on a non-flat field space background, we revisit the analytic expressions of the various cosmological observables such as scalar/tensor power spectra, scalar/tensor spectral tilts, non-Gaussianity parameters, tensor-to-scalar ratio, and the various runnings of these observables. In our backward formalism approach, the subsequent expressions of observables automatically include the terms beyond the leading order slow-roll expansion correcting many of the expression at subleading order.  To connect our analysis properly with the earlier results, we rederive the (well) known (single field) expressions in the limiting cases of our generic formulae. Further, in the light of PLANCK results, we examine for the compatibility of the consistency relations within the slow-roll regime of a two-field roulette poly-instanton inflation realized in the context of large volume scenarios.
\end{abstract}

\clearpage

%%%%%%%%%%%%%%%%%%%%%%%%%%%%%%%%%%%%%%%%%%%%%%%
%%%%%%%%%%%%%%%%%%%%%%%%%%%%%%%%%%%%%%%%%%%%%%%
%%%%%%%%%%%%%%%%%%%%%%%%%%%%%%%%%%%%%%%%%%%%%%%
%%%%%%%%%%%%%%%%%%%%%%%%%%%%%%%%%%%%%%%%%%%%%%%
%%%%%%%%%%%%%%%%%%%%%%%%%%%%%%%%%%%%%%%%%%%%%%%
%%%%%%%%%%%%%%%%%%%%%%%%%%%%%%%%%%%%%%%%%%%%%%%
%%%%%%%%%%%%%%%%%%%%%%%%%%%%%%%%%%%%%%%%%%%%%%%
%%%%%%%%%%%%%%%%%%%%%%%%%%%%%%%%%%%%%%%%%%%%%%%

\tableofcontents

\section{Introduction}

The inflationary paradigm has been proven to be quite fascinating for understanding various challenging issues (such as horizon problem, flatness problem, etc.) in the early universe cosmology \cite{Guth:1980zm,Linde:1981mu}. Moreover, it provides an elegant way for studying the inhomogeneities and anisotropies of the universe, which could be responsible for generating the correct amount of primordial density perturbations initiating the structure formation of the universe and the cosmic microwave background (CMB) anisotropies \cite{Planck:2013kta}. The simplest (single-field) inflationary process can be understood via a (single) scalar field slowly rolling towards its minimum in a nearly flat potential. %The vacuum fluctuations of the inflaton result in an almost scale invariant spectrum with a small tilt reflecting unique predictions via the CMB radiation \cite{Planck:2013kta}.
There has been enormous amount of progress towards constructing inflationary models and the same has resulted in plethora of those which fit well with the observational constraints from WMAP \cite{Larson:2010gs,Komatsu:2010fb} as well as the recent most data from PLANCK \cite{Planck:2013kta, Ade:2013zuv, Ade:2013ydc, Ade:2013uln}, and so far the experimental ingredients are not sufficient to discriminate among the various known models compatible with the experiments.

In general, if the perturbations are purely Gaussian, the statistical properties of the perturbations are entirely described by the two-point correlators of the curvature perturbations, namely the power spectrum. The observables which encode the non-Gaussian signatures are defined through the so-called non-linearity parameters $f_{NL}, \tau_{NL}$ and $g_{NL}$ parameter which are related to bispectrum (via the three-point correlators) and the tri-spectrum (via the four-point correlators) of the curvature perturbations. Although, the recent Planck data \cite{Ade:2013ydc} could not get very conclusive so far, it is still widely accepted that the signature of non-Gaussianity could be a crucial discriminator for the various known consistent inflationary models.
%This could be possibly detected in the upcoming results of PLANCK.
For this purpose, multi-field inflationary scenarios have been more promising because of their relatively rich structure and geometries involved \cite{Vernizzi:2006ve,Battefeld:2006sz,Choi:2007su,Rigopoulos:2005us,Seery:2006js,Byrnes:2009qy,Battefeld:2009ym} (See \cite{Byrnes:2010em,Suyama:2010uj} also for recent review). Meanwhile, a concisely analytic formula for computing the non-linear parameter for a given {\it generic } multi-field potential has been proposed in \cite{Yokoyama:2007dw,Yokoyama:2008by}, which is valid in the beyond slow-roll region as well. Recently, some examples with (non-)separable multifield potentials have been studied in \cite{Mazumdar:2012jj} which can produce large detectable values for the non-linear parameter $f_{NL}$ and $\tau_{NL}$. %Higher order correction to the non-linearity parameters have been studied in \cite{Lyth:2005fi,Zaballa:2006pv,Cogollo:2008bi,Rodriguez:2008hy}.
However, most of these works were investigated on a flat background. One of the main purpose of this work is to provide a general formula for these cosmological observables on a non-flat background in multi-filed inflationary model.

To illustrate the validity of these formula in a concrete model, we will utilize a so-called poly-instanton inflationary model which comes from the setup of string cosmology in Type IIB string compactification.
Significant amount of progress has been made in building up inflationary models in type IIB orientifold setups with the inflaton field identified as an open string modulus \cite{Kachru:2003sx,Dasgupta:2004dw,Avgoustidis:2006zp,Baumann:2009qx}, a closed string modulus \cite{Conlon:2005jm,Conlon:2008cj,Blumenhagen:2012ue} and involutively even/odd axions \cite{BlancoPillado:2004ns,Dimopoulos:2005ac,BlancoPillado:2006he,Kallosh:2007cc,Grimm:2007hs,Misra:2007cq,McAllister:2008hb}.
Along the lines of moduli getting lifted by  sub-dominant contributions, recently so-called poly-instanton corrections became  of interest. These are  sub-leading non-perturbative contributions which can be briefly described as instanton corrections to instanton actions. The mathematical structure of poly-instanton is studied in \cite{Blumenhagen:2012kz}, the consequent moduli stabilization and inflation have been studied in a series of papers \cite{Blumenhagen:2012ue,Cicoli:2011ct,Blumenhagen:2008kq,Lust:2013kt,Gao:2013hn}. %Meanwhile, the studies made in the context of axionic inflationary models in the type IIB orientifold framework have been quite promising too \cite{BlancoPillado:2004ns,Dimopoulos:2005ac,BlancoPillado:2006he,Kallosh:2007cc,Grimm:2007hs,Misra:2007cq,McAllister:2008hb}. In \cite{Burgess:2008ir}, inflation has been realized by a combination of the brane-motion and an axion while in \cite{Kallosh:2004rs,Bond:2006nc,BlancoPillado:2009nw}, it has been driven by a combination of a (divisor) volume mode and an axion.
In the framework of type IIB orientifolds, several single/multi-field models have been studied for aspects of non-Gaussianities \cite{Kallosh:2004rs,Burgess:2010bz,Misra:2008tx,Berglund:2010xr,Cicoli:2012cy, Gao:2013hn}. The computation of non-Gaussianties in racetrack models has been made in \cite{Sun:2006xv} and in the context of large volume scenarios, by
 the so-called roulette inflationary models \cite{Bond:2006nc,BlancoPillado:2009nw}. Despite of being a good and simple example for multi-field inflation with a non-flat background, this class of models allows the presence of several inflationary trajectories of sufficient ($\ge 50$) number of efoldings with significant curving and a subsequent investigation of non-Gaussianities in such a setup has resulted in small values of non-linearity parameters in slow roll \cite{Vincent:2008ds} and large detectable values of those in beyond slow-roll regime \cite{Gao:2013hn}.

%The tensor perturbations can be a source of the CMB temperature anisotropies produced during inflation, and the related signatures are described by another very important cosmological observable, namely the tensor-to-scalar ratio `$r$'.  Despite of encoding the information about the primordial gravitational waves, the tensor-to-scalar ratio $r$ can directly determine the inflationary energy scale and several interesting works have been done in this direction \cite{Lidsey:1995np, Byrnes:2006fr, Hotchkiss:2011gz, Choudhury:2013iaa, Li:2013nfa, Hebecker:2013zda, Li:2013nfa,Lello:2013awa}. A possible detection of $r$ in the near future experiments can help us in many ways to understand the inflationary physics deeper, and at the same time , it can serve as a discriminator for plethora of so far consistent inflationary models. Moreover, in \cite{Gong:2007ha}, it was motivated that running of tensor-to-scalar ratio $r$ could be relevant for detectability through laser interferometer experiments. On the similar lines, the importance of running of non-Gaussianiy parameters has also been motivated in \cite{Byrnes:2009pe, Byrnes:2012sc, Byrnes:2010ft, Leung:2012ve} and the same can be interesting in the light of the upcoming observations. See \cite{Suyama:2013nva,Byrnes:2013qjy} also, for a related analysis based on current Planck data.

In this article, our main aim is to revisit the analytic expressions of various cosmological observables, including scalar/tensor power spectra, scalar/tensor spectral tilts, non-Gaussianity parameters, tensor-to-scalar ratio and their runnings for a generic multi-field inflationary model driven on a non-flat background. The idea is to represent various observables in terms of field variations of the number of e-folding $N$ along with the inclusion of curvature correction coming from the non-flat field space metric. Some crucial developments along these lines have been made in recent works \cite{Yokoyama:2007dw, Byrnes:2012sc,Gong:2011uw, Elliston:2012ab, White:2013ufa, White:2012ya}. These generic expressions which automatically include the terms beyond the leading order slow-roll expansion, recover all the respective well known single field expressions in the limiting case. Moreover, we utilize these expressions for checking the various consistency relations in a string inspired two-field `roulette' inflationary model \cite{Gao:2013hn} based on poly-instanton effects. The strategy for computing the field-variations of number of e-folding $N$ is via numerical approach following  the so-called `backward formalism' \cite{Yokoyama:2007dw} and then to use the solutions for the computation of various cosmological observables. From the recent Planck data \cite{Planck:2013kta, Ade:2013zuv, Ade:2013ydc, Ade:2013uln}, the experimental bounds for various cosmological observables under consideration are,
\bea
\label{obs}
& & {\rm Scalar \, \, Power \, \, Spectrum: } \,  \, \, 2.092\times10^{-9} < {\cal P}_S < 2.297\times10^{-9} \nonumber\\
& & {\rm Spectral \, \, index:} \, \, \, 0.958  < n_S <  0.963 \nonumber\\
& & {\rm Running \, \, of \, \, spectral \, \, index:} \, \, \, -0.0098< \alpha_{n_S} <  0.0003\\
& & {\rm Tensor \, \, to \, \, scalar \, \, ratio:}  \, \, \, r < 0.11 \nonumber\\
& & {\rm Non\,\, Gaussianity \, \,parameters:} \, \, \, -9.8< f_{NL} < 14.3 , \, \, \tau_{NL} < 2800 \nonumber
\eea
while some other cosmological observables (like running of non-Gaussianity parameter) relevant for study made in this article could be important future observations.

The article is organized as follows: In section \ref{sec_Setup}, we will provide relevant pieces of information regarding type IIB orientifold compactification along with ingredients of  ``roulette-inflationary setup" developed with the inclusion of poly-instanton corrections \cite{Blumenhagen:2012ue, Gao:2013hn}. Section \ref{sec:ODEevolutionNabc} will be devoted to set the strategy for computing the field derivative of number of e-folding $N$ which gets heavily utilized in the upcoming sections. In section \ref{sec:cosmo-I}, we present the analytic expressions of various cosmological parameters such as scalar/tensor power spectra (${\cal P}_S, {\cal P}_T$), spectral index and tilt ($n_S, n_T$), tensor to scalar ratio ($r$) as well as their numerical details applied to the model under consideration. Section \ref{sec:cosmo-II} deals with a detailed analytical and numerical analysis of the non-linearity parameters ($f_{NL}, \tau_{NL}$ and $g_{NL}$) and their scale dependence encoded in terms of $n_{f_{NL}}, n_{\tau_{NL}}$ and $n_{g_{NL}}$ parameters.  Finally an overall conclusion will be presented in section \ref{sec_Conclusions and Discussions} followed by an appendix \ref{expressions} for intermediate computations.

\section{Roulette inflation setup with type IIB orientfolds}
\label{sec_Setup}
In order to illustrate the general formula for multi-field inflation model on a non-flat background,  we collect the relevant ingredients for a concrete model comes from type IIB orientifold compactification with the inclusion of poly-instanton corrections to the superpotential.   In the context of type IIB orientifolds compactification on Calabi-Yau threefolds $CY_3$,
it has been shown that in the presence of Wilson Divisor with $h^{1,0}_+(D)=1$, one has the right zero mode structure for an Euclidean D3-brane wrapping on it to generate poly-instanton effect in the superpotential \cite{Blumenhagen:2012kz}.

\begin{figure}[h!]
\centering
\includegraphics[scale=0.51]{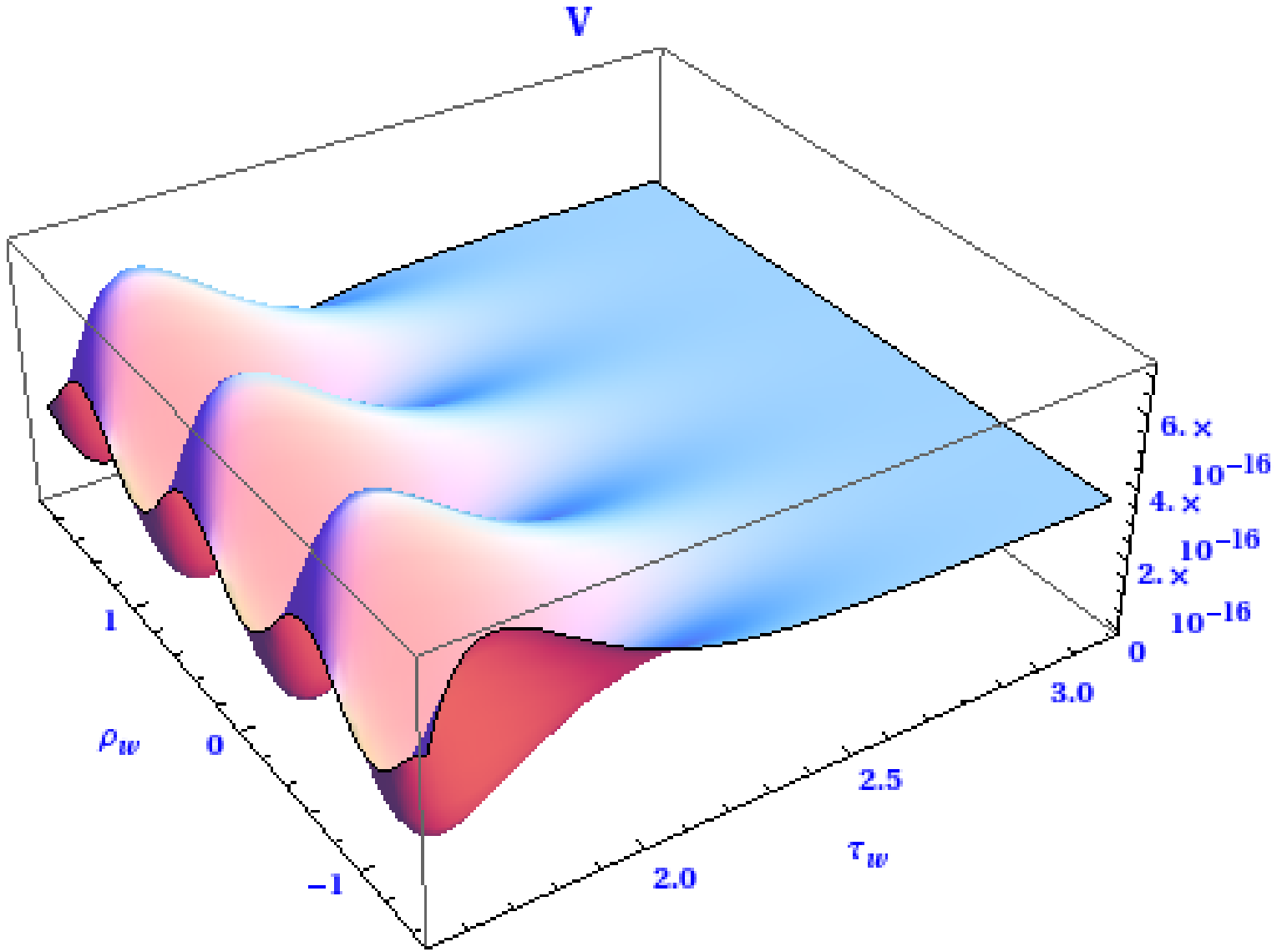}
\includegraphics[scale=0.51]{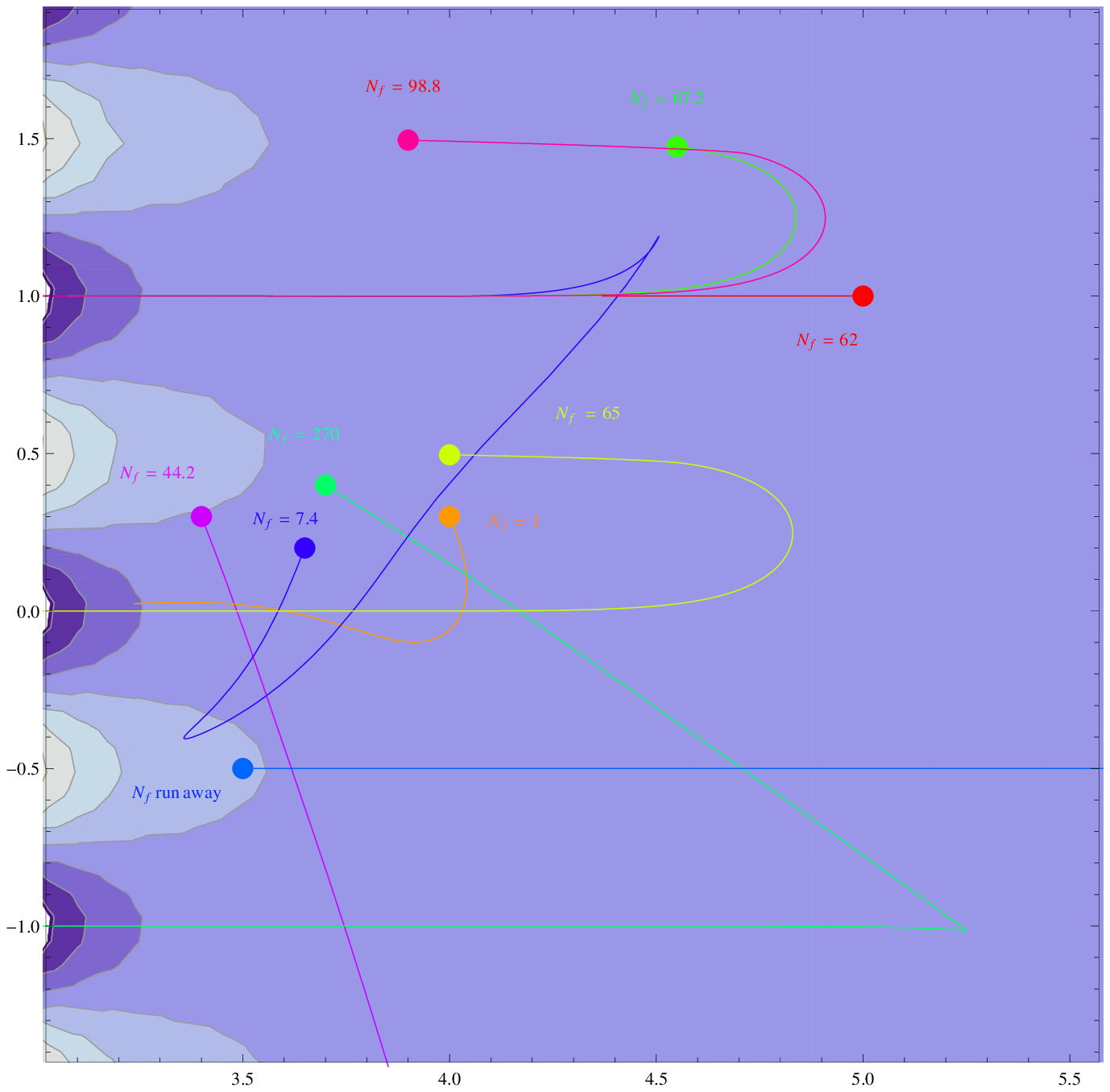}
\caption{The left figure is the effective potential as a function of the moduli $\tau_w$ and $\rho_w$. The right one is the  full inflationary trajectories for various initial conditions where the value of
e-folding number $N$  at the end of inflation is labeled on each of these trajectories.
 Various minima in dark blue are separated by maxima in light blue shade.
 }\label{potential}
\end{figure}

For $h^{11}_{-}(CY_3/{\cal O}) =0$, the ${\cal N} = 1$ K\"{a}hler
coordinates complexifying the four cycle volumes are simply given as $T_\alpha = \tau_\alpha + i \rho_\alpha$
\footnote{For a recent work related to implementing odd axion in poly-instanton setup and the relevant geometric configuration are studied in \cite{Gao:2013rra,Gao:2013pra}.}.
After  stabilizing the heavier moduli like the volume moduli $\cal V$ and small four-cicle moduli $T_s=\{\tau_s, \,\rho_s\}$ discussed in \cite{Blumenhagen:2012ue},
%we can get  the poly-Roulette inflation as have been done in \cite{Gao:2013hn}, i.e.
one gets a two-field potential of lighter moduli, i.e. the poly-instanton moduli $\tau_w$ and $\rho_w$,  which is simplified to the following expression after suitable uplifting mechanism \cite{Gao:2013hn}:
\bea
\label{effective_potential0}
& & \hskip-1.5cm V_{\rm inf}(\tau_w, \rho_w) = V_{\rm up} + V_0 + e^{-a_w \tau_w}\left(\mu_1 + \mu_2 \, \tau_w \right) \, \cos(a_w \rho_w) ~~ \,
\eea
where $a_w = 2 \pi$ while $\mu_1$, $\mu_2$, $V_0$ and $V_{\rm up}$ are model dependent parameters. This potential has the following set of critical points:
\bea
\ov\tau_w = \frac{\mu_2 - a_w \, \mu_1}{a_w \, \mu_2} , \, \, \,  a_w \ov\rho_w = m \, \pi
\eea
where $m\in \mathbb Z$. For the details of moduli stabilization and creating the mass hierarchy, we refers to the the reader to earlier work in \cite{Blumenhagen:2012ue, Gao:2013hn}. Moreover, in order to trust the effective field theory we need $\f{\mu_1}{\mu_2}<0$.
From now on, we fix our notation with a sampling of parameters such that $\{\mu_1 >0, \mu_2 < 0\}$ and performing the redefinitions $\tau_w = \phi_1, \, \rho_w = \phi_2$, the uplifted scalar potential becomes
\bea
\label{eq:Vinf}
& & \hskip-1.3cm V_{\rm inf} (\phi_1, \phi_2)=\left(\frac{g_s}{8 \pi}\right)\, e^{K_{\rm CS}} \, \left[-\frac{\mu_2 \, e^{-1+\frac{a_w \, \mu_1}{\mu_2}}}{a_w} + e^{-a_w \phi_1}\left(\mu_1 + \mu_2 \, \phi_1 \right) \, \cos(a_w \phi_2)\,\right].
\eea
Here, a proper normalization factor $\left(\frac{g_s}{8 \pi}\right)\, e^{K_{\rm CS}}$ has been included \cite{Conlon:2005jm}, where $K_{\rm CS}$ denotes the K\"ahler potential for the complex structure moduli. For the time being, we assume that $e^{K_{\rm CS}}\sim{\cal O}(1)$. Furthermore, we set the numerical parameters for moduli stabilization similar to the ones chosen in one of the benchmark models (in \cite{Blumenhagen:2012ue}). The parameters, which would be directly relevant for further computations in this article, are
\bea
\label{sampling}
 & & \mu_1 = 2.9\times10^{-8}, \, \mu_2 = -1.9\times 10^{-8}, \, a_w = 2 \pi, ~ g_s = 0.12 ~,~\\
& &  \hskip 2cm \ov{\cal V} = 905, \ov\tau_s = 5.7, \xi_{sw} = 1/(6 \sqrt2) ~.~\nonumber
\eea
The non-zero components of the `effective' non-flat moduli space metric ${\cal G}_{ab}$ relevant for inflaton dynamics are ${\cal G}_{11} \simeq \frac{3 \xi_{sw}}{2\,\sqrt2 \, \ov{\cal V}\, \sqrt{\ov{\tau_s}+{\phi_1}}} \simeq {\cal G}_{22}$. Note that the field space metric is diagonal and does not dependent on the second field $\phi_2$. The various non-zero  components of the Christoffel connections and the Riemann tensor are given as
\bea
& & \hskip-1cm \Gamma_{\,\, 11}^1 = \Gamma_{\, \, 12}^2 = - \Gamma_{\, \, 22}^1 \equiv -\frac{1}{4(\ov\tau_3 + \phi_1)}\, ; \, \, R_{\, \, 221}^1 = R_{\, \, 112}^2 \equiv \frac{1}{2(\ov\tau_3 + \phi_1)^2}. \nonumber
\eea
Under the sampling (\ref{sampling}), the form of the effective two-field inflationary potential (\ref{eq:Vinf}) is shown in Figure~\ref{potential} which leads to a ``roulette" type inflation \cite{Gao:2013hn}.

%\subsection{Roulette poly-instanton inflation}
%\label{sec_Evolution}
For the ``poly-roulette inflation'' proposed in \cite{Gao:2013hn},%the field evolutions and look for the possible inflationary trajectories which could reproduce 50 (or more) e-foldings. For a given initial condition, we numerically trace the entire inflationary trajectories including beyond slow-roll region.
we used the background $N$ e-folding number as the time coordinate, i.e. $dN = H dt$. The Einstein-Friedmann equations are obtained as
\begin{subequations}
\label{EOM}
\bea
\label{Friedmann}
\f{d^2}{dN^2}\phi^a+{\Gamma^a}_{bc}\f{d\phi^b}{dN}\f{d\phi^c}{dN}+\left(3+\f{1}{H}\f{dH}{dN}\right)\f{d\phi^a}{dN}+\f{{\cal G}^{ab} \partial_b V}{H^2}=0 ,
\eea
\bea
\label{constran}
H^2=\f{1}{3}\left(V(\phi^a)+\f{1}{2}H^2 \, {\cal G}_{ab} \f{d\phi^a}{dN}\f{d\phi^b}{dN} \right).
\eea
\end{subequations}
Using expressions (\ref{Friedmann}) and (\ref{constran}), one can derive another useful expression for variation of Hubble rate in terms of e-folding,
\bea
\label{third}
\f{1}{H}\f{dH}{dN}=\f{V}{H^2}-3 .
\eea
For numerical convenience, we solve these equations in the time basis $t$ and then change the result back to the basis $N$ e-folding.

As introduced in \cite{Yokoyama:2008by}, we will  follow the field redefinitions given as\footnote{The use of this notation would be more clear in the upcoming sections. Further, we will be using a combined indexing ${\cal A}$ such that any object ${\cal O}^{\cal A}$ has two components given as ${\cal O}^{\cal A}\equiv \{{\cal O}^a_1, {\cal O}^a_2\}$. For example, $\varphi^{\cal A}=\{\varphi^a_1, \varphi^a_2\}$.}
\bea
\label{eq:redef}
 \varphi^{a}_1\equiv\phi^a,  \,\, \varphi^{a}_2\equiv \dot{\phi}^a = \left(\f{d \phi^a}{d t}\right),
\eea
with $a=1,2$.
This redefinitions translate the second-order background equations of motions Eq.~(\ref{Friedmann}) into two first-order
Ordinary Differential Equations (ODEs) as follows
\bea
\label{EOM}
& & F^a_1 \equiv \frac{d\varphi^a_1}{dN}=\left(\frac{d \phi^a}{dN}\right)= \f{\varphi^a_2}{H} \, \, \, \, \, {\rm and} \, \, \, \, \,  F^a_2 \equiv \frac{D\varphi^a_2}{dN}= -3 \varphi^a_2 - {\cal G}^{ab} \f{V_b}{H} ~,~
\eea
where $D$ is the covariant derivative defined as $D \varphi^a_2 = d \varphi^a_2 + {\Gamma^a}_{bc} \varphi^b_2  d\varphi^c_1$ subject to the constraints
$$H^2=\f{1}{3}\left(V+\f{1}{2}\, {\cal G}_{ab} \varphi^a_2 \varphi^b_2 \right).$$ Then Eq.~(\ref{third}) will be simplified as $\dot H= -\f{1}{2}\, {\cal G}_{ab} \varphi^a_2 \varphi^b_2 $.
As usual, one has to look at the sufficient conditions for realizing slow-roll inflation which are encoded in the so-called slow-roll parameters
%For multi-field inflationary process with inflatons moving in a non-flat background, these slow-roll parameters are
$
 \epsilon \equiv -\f{\dot{H}}{H^2}, \,\,\,\,\,\,\,\, \eta \equiv \f{\dot{\epsilon}}{\epsilon H} .$
%& & \hskip-3.5cm {\rm and } \, \, \, \eta \equiv {\rm mostly \,\, negative \, \, eigenvalue \, \, of  \,\, the \, \,  matrix \, \,  \Upsilon_a^b \, \,  defined \, \, as} \nonumber\\
%& & \nonumber\\
%& &  \Upsilon_a^b = \frac{{\cal G}^{bc} \, \left(\Gamma^d_{ac} \, \partial_d V_{\rm inf}-  \partial_{ac} V_{\rm inf}\right)}{V_{\rm inf}}
 Now, we can solve the background field equations (\ref{EOM}) to get the full trajectories under different initial conditions. We choose  $\phi^a(0)=\phi^a_0 \, \, {\rm and} \, \, \f{d\phi^a}{dt}\f{d\phi_a}{dt}|_{t=0}=0;\, \, {\rm for} \, \,  a \in \{1,2\}$
as  a set of initial conditions and trace the corresponding trajectories up to the end of inflation. Figure \ref{potential} shows the complex evolution of trajectories for some samples of initial conditions given in Table \ref{initial}.
\begin{table}[H]
  \centering
  \begin{tabular}{|c||c|c||c|c|}
  \hline
    Class & $\tau_w$  & $\rho_w$  & $N_F$ & Trajectory    \\
    \hline \hline
     I  &  5 & 1 & 62   &  I \\
  \hline
   & 4 & 0.3 & 1 & \\
  II   & 4.55 & 1.474 & 67.2  &  \\
   & 4 & 0.496 & 65 &  IIa\\
     & 3.9 & 1.495 & 98.8 & IIb \\
     \hline
     III  & 3.5 & -0.5 & - & \\
    \hline
    & 3.65 & 0.2 & 7.4 &  \\
 IV & 3.4 & 0.3 & 44.2 & \\
      & 3.7 & 0.4 & 270 & IV\\
    \hline
  \end{tabular}
  \caption{Initial conditions for these trajectories shown in Figure~\ref{potential}. The trajectories I, IIa, IIb and IV are chosen for studying cosmological observables in the upcoming sections.}
  \label{initial}
 \end{table}
The various inflationary trajectories shown in Figure~\ref{potential} can be classified in the following categories
\begin{enumerate}[(I)]
\item Given that the initial conditions are such that the axion is minimized at its respective minimum to begin with, two-field inflationary process reduces to its single field analogue  \cite{Blumenhagen:2012ue}. These are stable trajectories and are attracted towards the respective valley in a straight line like the trajectory in Figure~\ref{potential} with $N_F=62$.
%These can produce the required number of e-foldings if the Wilson divisor volume mode is displaced significantly away from the minimum.
\item For the axion initial condition being a little bit (and not too far) away from the minimum, the trajectories rolls to the nearest valley and trace towards the respective minimum like those trajectories in Figure~\ref{potential} with $N_F=1, \, 67.2, \, 65, \, 98.8$.
\item If the axion initial condition starts with its value at the maximum, this results in an unstable trajectory directed straightly outwards from the respective attractor point showing a run-away behavior like the yellow trajectory in Figure~\ref{potential}.
\item If a trajectory starts from the axion initial condition being closer to some maximum value as well as the initial value for the divisor volume mode being not very far from its respective minimum, one observes that such an inflationary trajectory crosses several axion-ridges before getting attracted into a valley. This can be understood from the fact that this class of initial condition is such that the initial potential energy is just a little higher to begin with and the $N$ e-folding increase very slow at the beginning of these trajectories, see Figure~\ref{potential} with $N_F=7.4,\, 44.2, \, 270$.
\end{enumerate}

For most of these trajectories except the single-field one, there exists a region of quick-roll (with $\eta >1$) before starting the slow-roll. However, this region lasts within a couple of e-foldings.  Further, there is a region in field space where there is a {\it strong} violation of slow-roll condition via $\eta \gg 1$ before the end of inflation. This beyond slow-roll regime also does not significantly contribute to the e-folding and lasts within one or two e-foldings. In this article, our main focus has been to look for the behavior of various cosmological parameters within the slow-roll regime which covers the most of the inflationary process.

\section{Field derivatives of number of e-foldings ($N$)}
\label{sec:ODEevolutionNabc}
The field variations of number of e-foldings  (denoted as $N_{{\cal A}_1{\cal A}_2....{\cal A}_n}$) play very crucial role as most of the cosmological observables can be written out by utilizing the same, and hence computing those is always among the central task.
%To begin with, let us briefly start with stating the $\delta N$-formalism which will be useful throughout this section.  Our expression generalize the previous works \cite{Yokoyama:2007dw,Yokoyama:2008by} in a general non-flat background up to cubic order. Choosing a proper gauge \cite{Sasaki:1998ug}, one can  treat the number of e-folding $N$ as time coordinate. Then one can relates the curvature perturbations to the difference between the number of e-foldings $\delta N$ of two constant time-hypersurfaces \cite{Maldacena:2002vr},
%\bea
%\label{eq:deltaNdef}
%\zeta(t,x) \simeq \delta N = H \delta t ~.~
%\eea
Following \cite{Yokoyama:2007dw,Yokoyama:2008by} on the lines of the redefinitions (\ref{eq:redef}) in the previous section, the perturbations of the scalar field on $N = {\rm constant}$ gauge can be expressed as
\bea
\delta \varphi^{\cal A}(\lambda, N) \equiv \varphi^{\cal A}(\lambda + \delta\lambda, N)-\varphi^{\cal A}(\lambda, N)~,~
\eea
where $\lambda$'s are $2 {n}-1$ integration constants (for an $n$- component scalar field) which, along with $N$, parametrizes the initial values of the fields \cite{Yokoyama:2007dw,Yokoyama:2008by}.
Further, considering the field fluctuations in $N=$ const. gauge, the $\delta N$ formalism \cite{Maldacena:2002vr} implies expressing the curvature perturbations at each spatial point of the field space at $N=N_F$ where $N_F$ corresponds to a final time-hypersurface of uniform energy density. In fact, the curvature perturbations can be expressed at each spatial point in terms of the variation of the field fluctuations point to point and order by order as under \cite{Yokoyama:2008by}
\bea
\label{eq:deltaNgen2}
& & \hskip-1.7cm \zeta(N_F, {\bf x}) \simeq
\delta N (N_F, \varphi^{\cal A}(N^*))\nonumber\\
& & =\sum\frac{1}{n !} N^*_{{\cal A}_1{\cal A}_2....{\cal A}_n} \, \delta\varphi^{{\cal A}_1}({\bf x}) \, \delta \varphi^{{\cal A}_2}({\bf x}) ......\delta \varphi^{{\cal A}_n}({\bf x}),\\
& & \hskip-0.3cm N^*_{{\cal A}_1{\cal A}_2....{\cal A}_n} \equiv \left(\frac{D^n N(N_F, {\varphi^{\cal A}})}{\partial\varphi^{{\cal A}_1}\partial\varphi^{{\cal A}_2}....\partial\varphi^{{\cal A}_n}}\right)_{at \, \, \, \varphi^{\cal A} = \varphi^{\cal A}_{(0)}(N_*)}
~,~\nonumber
\eea
where $\varphi^{\cal A}_{(0)}$ corresponds to an unperturbed trajectory and  the quantities with superscripts $*$ mean to be evaluated at the initial time $N=N_*$. Moreover, due to spatial dependence, the values of fields $\varphi^{\cal A}_0(N_*)$ on the initial flat hypersurface differ point to point and thus characterizes the initial field perturbations. As the number of e-foldings is counted between the initial and final hypersurface, it has field (${\varphi}^{\cal A}$) dependence in terms of the fluctuation vector $\delta\phi^{\cal A}$ as given under
\bea
\label{eq:evo1}
& & \delta\varphi^{\cal A} \equiv \varphi^{\cal A} -  \varphi^{\cal A}_0=  \delta^{(1)} \varphi^{\cal A} + \frac{1}{2} \, \delta^{(2)} \varphi^{\cal A} +  ........ \, \, ,
\eea
Also, by $\delta N$ formalism, the field derivatives of the e-folding $N^*_{{\cal A}_1{\cal A}_2....{\cal A}_n}$ are simply given by field derivatives of $N(N_F, \varphi^{\cal A})$ which being the number of e-folding gained during the evolution of the homogeneous universe from an initial to a final uniform energy density hypersurface, and hence field variations $N^*_{{\cal A}_1{\cal A}_2....{\cal A}_n}$ will also have dependence on the number of e-foldings through $\varphi_{\cal A}$. Now we come to the task of computing these field derivatives of e-foldings which is important for cosmological observable computations.

The dynamics of the field derivative of e-foldings can be expressed in terms of coupled first order differential equations. To establish those relations, the evolution equations for the field fluctuations $\delta \varphi^{\cal A}$ is an important ingredient. The same can be obtained by perturbing the dynamical equation (\ref{EOM}) for a non-flat background metric, and are simply given by order by order as under\cite{Yokoyama:2008by}
\bea
\label{eq:evolution1}
& & \frac{D}{dN} \, \delta \varphi^{\cal A}(N) = P^{\cal A}_{\, \, \, \, \cal B}(N) \, \delta \varphi^{\cal B}(N)+ \frac{1}{2} \, Q^{\cal A}_{(3) \, \, \, \cal B \cal C}(N) \, \delta \varphi^{\cal B}(N) \, \delta \varphi^{\cal C}(N) + ......\\
& & \hskip2cm  + \frac{1}{(l-1)!} \, \, Q^{\cal A}_{(l) \, \, \, {\cal B}_1 ....{\cal B}_{l-1}}(N) \, \delta \varphi^{{\cal B}_1}(N)...... \, \delta \varphi^{{\cal B}_{l-1}}(N) \, + ...... ~,~\nonumber
\eea
where $P^{\cal A}_{\, \, \, \, \cal B}$ and $Q^{\cal A}_{(l) \, \, \, {\cal B}_1 ....{\cal B}_{l-1}}$ are defined as follows:
\bea
\label{eq:PandQdef}
& & P^{\cal A}_{\, \, \, \, \cal B} \equiv \left(\frac{D F^{\cal A}}{\partial \varphi^{\cal B}}\right)_{at \, \, \, \varphi^{\cal A} = \varphi^{\cal A}_{(0)}(N)}~,~\\
& & Q^{\cal A}_{(l) \, \, \, {\cal B}_1 ....{\cal B}_{l-1}} \equiv  \left(\frac{D^{l-1}\, F^{\cal A}}{\partial \varphi^{{\cal B}_1}\, \partial \varphi^{{\cal B}_2}.......\partial \varphi^{{\cal B}_{l-1}}}\right)_{at \, \, \, \varphi^{\cal A} = \varphi^{\cal A}_{(0)}(N)}
~,~\nonumber
\eea
where $\varphi^{\cal A}_{(0)}$ corresponds to an unperturbed trajectory. For example, using the dynamics of fields $\varphi^{\cal A}$ governed by  (\ref{EOM}), the explicit expressions for $P^{\cal A}_{\,\,\, {\cal B}}(N)$ are simplified to
\bea
\label{eq:Pexplicit}
& & P^{a1}_{\, \, \, 1b}= -\frac{1}{6 H^3} \, \varphi^a_2 \, V_b ~,~\nonumber\\
& & P^{a1}_{\, \, \, 2b}= -\frac{V^a_{\, \, \, b}}{ H}+ \frac{1}{6 H^3} \, V^a V_b -\frac{1}{H}\,R^a_{\, \,  \,c b d} \, \varphi^c_2 \, \varphi^d_2~,~\\
& & P^{a2}_{\, \, \, 1b}= \frac{1}{H} \delta^a_b -\frac{1}{6 H^3} \, \varphi^a_2 \, ({\cal G}_{bd}\varphi^d_2) ~,~\nonumber\\
& & P^{a2}_{\, \, \, 2b}= -3 \, \delta^a_b + \frac{1}{6 H^3} \, V^a \, ({\cal G}_{bc}\varphi^c_2)~. \nonumber
\eea
The other expressions for $Q^{\cal A}_{(l) \, \, \, {\cal B}_1 ....{\cal B}_{l-1}}$ can be analogously computed by using the higher order covariant field $\varphi^{\cal A}$ derivatives of $F^{\cal A}$. Now consider the variations of curvature perturbation defined in (\ref{eq:deltaNgen2}) as under
\bea
& & \hskip-1.8cm \frac{D}{dN} \, \zeta(N) = \biggl[\left(\frac{DN_{\cal A}}{dN}\right) \, \delta\phi^{\cal A} + N_{\cal A} \, \left(\frac{D\delta\phi^{\cal A}}{dN}\right) \biggr]  \\
& & + \frac{1}{2 !} \biggl[ \left(\frac{DN_{\cal A \cal B}}{dN}\right) \, \delta\phi^{\cal A} \,\, \delta\phi^{\cal B}  + N_{\cal A \cal B} \, \left(\frac{D\delta\phi^{\cal A}}{dN}\right) \,\, \delta\phi^{\cal B} + N_{\cal A \cal B}  \,\, \delta\phi^{\cal A} \, \left(\frac{D\delta\phi^{\cal B}}{dN}\right) \biggr]  \nonumber\\
& & + \frac{1}{3 !} \biggl[ \left(\frac{DN_{\cal A \cal B \cal C}}{dN}\right) \, \delta\phi^{\cal A} \,\, \delta\phi^{\cal B}\,\, \delta\phi^{\cal C}  + N_{\cal A \cal B \cal C} \, \left(\frac{D\delta\phi^{\cal A}}{dN}\right) \,\, \delta\phi^{\cal B} \,\, \delta\phi^{\cal C} \nonumber\\
& & + N_{\cal A \cal B \cal C} \,\, \delta\phi^{\cal A} \, \left(\frac{D\delta\phi^{\cal B}}{dN}\right) \,\, \delta\phi^{\cal C} + N_{\cal A \cal B \cal C} \,\, \delta\phi^{\cal A} \,\, \delta\phi^{\cal B} \, \left(\frac{D\delta\phi^{\cal C}}{dN}\right)  \biggr] + .......\nonumber
\eea
Using the expressions (\ref{eq:evo1}), and the fact that curvature perturbation at final uniform hypersurface $N_F$ is independent on the choice of $N_F$ as long as $N_F > N_c$, where $N_c$ is certain time after background trajectories have completely conversed, then in a backward evolution manner,  the {\it constancy} of curvature perturbation at $N =N_F$ can be ensured order by order by satisfying the following backward evolution differential equations given as under~\footnote{Expressions analogous to (\ref{eq:ODEsfieldN}) can also be found in \cite{Elliston:2012ab}. Although our strategy (which is based on backward-formalism) is the same to those of \cite{Yokoyama:2007dw,Yokoyama:2008by}, however our approach for solving the ODEs is different as for our case the sole task has been reduced to solve coupled ODEs of tensorial objects ($N_{\cal A}, N_{\cal A \cal B}$ and $N_{\cal A \cal B \cal C}$) instead of vector objects ($N_{\cal A}, \Theta^{\cal A}$ and $\Omega_{\cal A}$) as in \cite{Yokoyama:2007dw,Yokoyama:2008by}. See \cite{Yokoyama:2007uu} also for similar computation on {\it constancy} of curvature perturbation at $N=N_F$.}
\bea
\label{eq:ODEsfieldN}
& & \frac{D}{dN} { N_{\cal A}}(N) = - \,  { N_{\cal B}(N)\,  P^{\cal B}_{\, \, \, {\cal A}}|_{\varphi=\varphi^{(0)}(N)}} \, ,\\
& & \hskip-0.2cm\frac{D}{dN} { N_{\cal A \cal B}}(N)  = - N_{\cal A \cal C}(N) \, P^{\cal C}_{\, \, \, {\cal B}} - N_{\cal B \cal C}(N) \, P^{\cal C}_{\, \, \, {\cal A}}  - N_{\cal C}(N) \, Q^{\cal C}_{\, \, \, \cal A \, {\cal B}}|_{\varphi=\varphi^{(0)}(N)}\, , \nonumber\\
& & \hskip-0.4cm \frac{D}{dN} { N_{\cal A \cal B \cal C}}(N)  = -\, N_{\cal D}(N) \, Q^{\cal D}_{\, \,  \, {\cal A} \, \cal B \, \cal C} \, - N_{\cal A \cal B \cal D}(N) \, P^{\cal D}_{\, \, \, \cal C} - N_{\cal A \cal D \cal C}(N) \, P^{\cal D}_{\, \, \, \cal B} - N_{\cal C \cal B \cal D}(N) \, P^{\cal D}_{\, \, \, \cal A} \nonumber\\
& & \hskip2.5cm - \, N_{\cal C \, {\cal D}}(N) \, Q^{\cal D}_{\, \,  \, {\cal A} \, \cal B}\,- \, N_{\cal A \, {\cal D}}(N) \, Q^{\cal D}_{\, \,  \, {\cal B} \, \cal C}\,-   \,N_{\cal B \, {\cal D}}(N) \, Q^{\cal D}_{\, \,  \, {\cal C} \, \cal A} |_{\varphi=\varphi^{(0)}(N)}\nonumber
\eea
where it is understood that all the quantities in the right hand side of the aforementioned expressions depend on e-folding number $N$. The initial conditions for solving the above set of ODEs, which are the values of various derivatives of e-folding $N$ evaluated at some final constant time-hypersurface $t_F$ (e.g. $N_{\cal A}^F, N_{{\cal A}{\cal B}}^F, N_{{\cal A}{\cal B}{\cal C}}^F$), are given as follows
\bea
\label{eq:Bds}
& & N_{\cal A}^F = -\left(\frac{H_{\cal A}}{H_{\cal D}\, F^{\cal D}}\right)_{at \, \, \varphi=\varphi^{(0)}(N_F)} ~,~\nonumber\\
& & N_{{\cal A}{\cal B}}^F = -\left(\frac{U_{{\cal A}{\cal B}}}{H_{\cal D}\, F^{\cal D}}\right)_{at \, \, \varphi=\varphi^{(0)}(N_F)}~,~\\
& & N_{{\cal A}{\cal B}{\cal C}}^F = -\left(\frac{Z_{{\cal A}{\cal B}{\cal C}}}{H_{\cal D}\, F^{\cal D}}\right)_{at \, \, \varphi=\varphi^{(0)}(N_F)}~.~\nonumber
\eea
The expressions for quantities $H_{\cal A}(N), H_{{\cal A}{\cal B}}(N), H_{{\cal A}{\cal B}{\cal C}}(N)$,$ U_{{\cal A}{\cal B}}(N)$,${ Z_{{\cal A} {\cal B}{\cal C}}}(N)$ as well as ${ Q^{\cal A}_{\,\,\,\, {\cal B}{\cal C}}}(N)$ and ${ {\cal Q}^{\cal A}_{\, \, \, {\cal B}{\cal C}{\cal D}}}(N)$ involve various derivatives of the scalar potential and the Hubble rate. The explicit expressions can be found in Appendix of \cite{Gao:2013hn}.

%one can separate out the various pieces of (\ref{eq:evolution1}) order by order as under,
%\bea
%\label{eq:evo2}
%& & \frac{D}{dN}\, \delta^{(1)} \phi^{\cal A} = P^{\cal A}_{\, \, \, {\cal B}} \, \, \delta^{(1)} \phi^{\cal B} \\
%& & \frac{D}{dN}\, \delta^{(2)} \phi^{\cal A} = P^{\cal A}_{\, \, \, {\cal B}} \, \, \delta^{(2)} \phi^{\cal B} + Q^{\cal A}_{\, \, \, {\cal B}{\cal C}} \, \, \delta^{(1)} \phi^{\cal B} \, \, \delta^{(1)} \phi^{\cal C} \nonumber\\
%& & \frac{D}{dN}\, \delta^{(3)} \phi^{\cal A} = P^{\cal A}_{\, \, \, {\cal B}} \, \, \delta^{(3)} \phi^{\cal B} + \frac{3}{2} \, Q^{\cal A}_{\, \, \, {\cal B}{\cal C}} \, \, \delta^{(2)} \phi^{\cal B} \, \, \delta^{(1)} \phi^{\cal C} +\frac{3}{2} \, Q^{\cal A}_{\, \, \, {\cal B}{\cal C}} \, \, \delta^{(1)} \phi^{\cal B} \, \, \delta^{(2)} \phi^{\cal C} \nonumber\\
%& & \hskip3cm +  \, Q^{\cal A}_{\, \, \, {\cal B}{\cal C}\cal D} \, \, \delta^{(1)} \phi^{\cal B} \, \, \delta^{(1)} \phi^{\cal C}\, \, \delta^{(1)} \phi^{\cal D}\nonumber
%\eea

In our two field model described in previous section, the set of equations (\ref{eq:ODEsfieldN}) expands into 84 (4 + 16 + 64) coupled differential equations which have to be numerically solved utilizing the same number of conditions given in (\ref{eq:Bds}). After having the numerical solutions to these field derivatives, one can easily compute all the cosmological observables as the same can be written in terms of $N_{\cal A}, ~N_{\cal A \cal B}$ and $N_{\cal A \cal B \cal C}$. In the upcoming section we would revisit the generic analytic expressions for the various cosmological observables and subsequently analyze the numerical estimates.

\subsection*{Various expressions for a single field inflationary potential}

In order to make our notations sufficiently clear and convenient to follow, let us briefly present
the simplified version of those expressions for a single field inflationary potential $V(\phi)$
driven on a flat background. {\it The same would be useful to derive the well-known single field
expressions for cosmological observables} such as scalar power spectrum $P_s$, spectral index $n_s$,
running of spectral index $\alpha_{n_s}$ etc., whose general multi-field forms for
a non-flat background have to be discussed later in the upcoming sections.

The generalized two-component vector $\phi^{\cal A} = \{ \phi^a_1, \, \phi^a_2\}$ is simply given as $\phi^{\cal A} = \{ \phi, \, \dot\phi \}$. The inflaton dynamics is governed by the second order EOM given as $\ddot\phi + 3\, H \, \dot\phi + V_\phi = 0$ which can be reformulated into two first-order expressions as under
\bea
\label{eq:single1}
& & F^{\cal A}:= \, \, \, \, \, \, \, \, \, \, F^{\phi} = \frac{\dot\phi}{H} \, \, , \, \, \, \, \, F^{\dot\phi} = - 3 \, \dot\phi - \frac{V_\phi}{H}.
\eea
Using these expressions of $F^{\cal A}$, the simplified versions of various components of $P^{\cal A}_{\, \, \, \, {\cal B}}$ are written as
\bea
\label{eq:single2}
%& & N_{A} \equiv \{N_a^1 , N_a^2\} = \{N_\phi, N_{\dot \phi}\}  ~,~\\
& & P^{\phi}_{\, \, \, \phi}= -\frac{1}{6 H^3} \, \dot \phi \, V_\phi \, , \, \, \, \, \, P^{\dot \phi}_{\, \, \, \phi}= -\frac{V_{\phi \phi}}{ H}+ \frac{1}{6 H^3} \, V_\phi V_\phi  ~,~\\
& & P^{\phi}_{\, \, \,\dot \phi}= \frac{1}{H} -\frac{1}{6 H^3} \, {\dot \phi}^2 \, , \, \, \, \, \, P^{\dot \phi}_{\, \, \, \dot \phi}= -3 \,  + \frac{1}{6 H^3} \, V_\phi \, \dot \phi  ~.~\nonumber
\eea
Similarly, the eight components of ${Q_{(3)}}^{\cal A}_{\, \, \, \, {\cal B} {\cal C}}$ are simplified to
\bea
\label{eq:single3}
%& & N_{A} \equiv \{N_a^1 , N_a^2\} = \{N_\phi, N_{\dot \phi}\}  ~,~\\
& & Q^{\phi}_{\, \, \, \phi \phi}=  \frac{V_\phi^2 \, \dot\phi}{12 \, H^5} - \frac{V_{\phi\phi} \, \dot\phi}{6 \, H^3} \, , \, \, \, \, \, Q^{\dot\phi}_{\, \, \, \phi \dot\phi}=  -\frac{V_\phi^2 \, \dot\phi}{12 \, H^5} + \frac{V_{\phi\phi} \, \dot\phi}{6 \, H^3}   ~,~\nonumber\\
& & Q^{\phi}_{\, \, \,\dot\phi \phi}= -\frac{V_\phi}{6 \, H^3} + \frac{V_\phi \, \dot\phi^2}{12 \, H^5} \, , \, \, \, \, \, Q^{\phi}_{\, \, \,\dot\phi \dot\phi}= -\frac{\dot\phi}{2 \, H^3} + \frac{\dot\phi^3}{12 \, H^5}  ~,~\\
& &Q^{\dot\phi}_{\, \, \, \phi \phi}=  -\frac{V_\phi^3}{12 \, H^5} + \frac{V_\phi \, V_{\phi \phi}}{2 \, H^3} - \frac{V_{\phi \phi \phi}}{H} \, , \, \, \, \, \,  Q^{\dot\phi}_{\, \, \, \dot\phi \phi}=  -\frac{V_\phi^2 \, \dot\phi}{12 \, H^5} + \frac{V_{\phi\phi} \, \dot\phi}{6 \, H^3}  ~,~\nonumber\\
& &  Q^{\phi}_{\, \, \, \phi \dot\phi}= -\frac{V_\phi}{6 \, H^3} + \frac{V_\phi \, \dot\phi^2}{12 \, H^5}  \, , \, \, \, \, \, Q^{\dot\phi}_{\, \, \, \dot\phi \dot\phi}=  -\frac{V_\phi \, \dot\phi^2}{12 \, H^5} + \frac{V_{\phi}}{6 \, H^3} ~,~\nonumber
\eea
while the sixteen components of ${{\cal Q}_{(4)}}^{\cal A}_{\, \, \, \, {\cal B}\, {\cal C} \, {\cal D}}$
are given in the appendix \ref{sec_QABCD}.

\section{Cosmological observables-I}
\label{sec:cosmo-I}
\subsection{Scalar power spectra, spectral index  and its scale dependence}
\subsubsection*{Scalar power spectrum ($\cal P_S$)}
Utilizing the generalized field derivatives of the number of e-foldings $N$, power spectra of the scalar perturbation modes for a multi-field inflation driven on a non-flat background can be simply given as \cite{Yokoyama:2008by}
\bea
\label{eq:Ps}
& & {\cal P}_S = \left(\frac{H^2}{4 \, \pi^2} \, {A}^{{\cal A} {\cal B}} \, N_{\cal A} \,
N_{\cal B}\right)_{at \, \, \, N = N_*} ~,~
\eea
where the field variations of $N$ are defined as $N_{\cal A}=D_{\cal A}N, N_{{\cal A}{\cal B}} = D_{{\cal A}{\cal B}} N$ and $N^{\cal A} = {A}^{{\cal A}{\cal B}} N_{\cal B}$. %Here, the symbol $ {A^{{\cal A}{\cal B}}}$ is defined through the following correlator of two field fluctuations $\delta \varphi^{\cal A}$
%\bea
%& &  \left<\delta \varphi^{\cal A}_* \, \delta \varphi^{\cal B}_*\right> = { A^{{\cal A}{\cal B}}}\, \left(\frac{H_*}{2\pi}\right)^2 .
%\eea
In general, $ {A^{{\cal A}{\cal B}}}$ depends on the non-flat background metric. The explicit expressions for various components, after including the slow-roll corrections \cite{Nakamura:1996da, Byrnes:2006vq, Byrnes:2006fr}, are given in Appendix~\ref{expressions}. Now, after expanding the various terms in (\ref{eq:Ps}), we get
\bea
\label{eq:Ps0}
& & {\cal P}_S = \frac{H^2}{4 \, \pi^2} \, \biggl[{A^{ab}_{11}} \, N_a^1 \, N_b^1 + {A^{ab}_{12}} \, N_a^1 \, N_b^2 +{A^{ab}_{21}} \, N_a^2 \, N_b^1+ {A^{ab}_{22}} \, N_a^2 \, N_b^2\biggr]\\
& & = \biggl[\frac{H^2}{4 \, \pi^2} \, {A^{ab}_{11}} \, N_a^1 \, N_b^1 \biggr]+\biggl[\frac{H^2}{4 \, \pi^2} \,  \left( {A^{ab}_{12}} \, N_a^1 \, N_b^2 +{A^{ab}_{21}} \, N_a^2 \, N_b^1\right)\biggr]+ \biggl[\frac{H^2}{4 \, \pi^2} \,  {A^{ab}_{22}} \, N_a^2 \, N_b^2\biggr]\nonumber\\
& &   \hskip 2cm I \, \, \, \hskip 4.5cm II \, \, \, \hskip 4cm III ~~\nonumber
\eea
\begin{figure}[h!]
\centering
\includegraphics[scale=0.95]{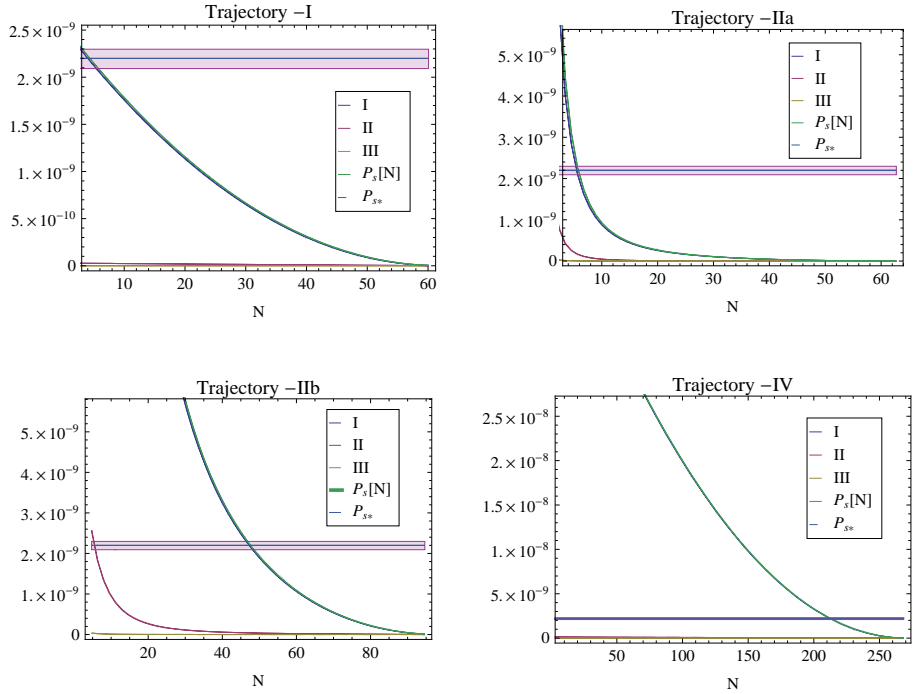}
\caption{Scalar power spectrum plotted for the four trajectories under consideration. It is observed that the most dominant contribution comes from terms of type-I as mentioned in (\ref{eq:Ps0}). The shadow region shows the allowed window (\ref{obs}) presented by Planck \cite{Planck:2013kta, Ade:2013zuv, Ade:2013ydc, Ade:2013uln}.}
\label{Ps}
\end{figure}
In Figure \ref{Ps}, the blue lines inside the shadow represents an intermediate value ($P_{s*} \sim 2.1 \times 10^{-9}$) allowed in the constraint window. Depending on the hierarchal contributions expected\footnote{Please see Appendix~\ref{expressions} for details on components of $A^{\cal A \cal B}$ and a numerical justification about the slow-roll relation $3 H \, N_a^2 \sim N_a^1$.} from the metric components $A^{\cal A \cal B}$, we separate out the respective three kinds of terms in (\ref{eq:Ps0}) for numerical investigations. A numerical analysis as shown in Figure \ref{Ps} confirms that the most dominant contribution comes from the first piece (I) of Eq.~(\ref{eq:Ps0}). The first piece-I, which produces almost entire scalar power spectrum ${\cal P}_S$, can also be rewritten as\footnote{Please see the appendix \ref{expressions} for the details.}
\bea
\label{eq:Ps0new}
& & {\cal P}_S=  \left(\frac{H_*}{2\pi}\right)^2 \, \left[{\cal G}^{ab} - 2 \, \epsilon \, \, {\cal G}^{ab} + 2 \, \alpha \, \frac{{\cal G}^{ac} \epsilon_{cd} N^d_1 \, N^b_1}{{\cal G}^{pq} \, N_p^1\, \, N_q^1}\right] \, N_a^1 \, N_b^1 ~,~
\eea
where in the above expression, $\alpha = 2 - \ln 2 - \gamma \simeq 0.7296$ with $\gamma \simeq 0.5772$  the Euler-Mascheroni constant \cite{Nakamura:1996da, Byrnes:2006vq, Byrnes:2006fr}, and $\epsilon_{ab}$ is defined as
$$ \epsilon_{ab} = \epsilon \, {\cal G}_{ab} + \left({\cal G}_{ac} \, {\cal G}_{bd} -\frac{1}{3} \, R_{abcd}\right) \, \frac{\varphi^c_2 \, \varphi^d_2}{H^2} - \frac{V_{;ab}}{3 \, H^2}~.~$$
Further, for a single field ($\phi$) inflationary model, using the slow-roll relations $N^2_{a}\equiv N_{\dot\phi} \sim \frac{N_{\phi}}{3\,H}$ along with the simplified definitions $N^1_{a}\equiv N_{\phi} = \frac{H}{\dot\phi}$ and $\epsilon = \frac{\dot\phi^2}{2 \, H^2} $, we get a simple and well-known result~\cite{Byrnes:2006fr,Sasaki:1995aw, Lidsey:1995np}
\bea
\label{eq:Ps0new1}
{\cal P}_S \sim \frac{H^2}{4 \, \pi^2} \, \biggl[{\cal G}^{ab}  \, N_a^1 \, N_b^1\biggr]\sim \frac{H^2}{4 \, \pi^2 \, (2 \, \epsilon)}~.~\,
\eea
{Apart from recovering the well known expressions (\ref{eq:Ps0new}) (as given in \cite{Nakamura:1996da, Byrnes:2006vq}) and (\ref{eq:Ps0new1})  (as can be found in \cite{Byrnes:2006fr,Sasaki:1995aw, Lidsey:1995np}) in the limiting cases, our general expression (\ref{eq:Ps0}) for scalar power spectrum involves new contributions; for example, the second (II) and third pieces (III) of (\ref{eq:Ps0}) are new terms in our analysis which includes the contributions of the types involving the derivatives of generalized (twofold) field vector $\varphi^{\cal A} \equiv \{\varphi^a_1, \varphi^a_2\}$, i.e. not only the field vector $\phi^a = \varphi^a_1$ but also the derivatives of the time-derivatives of the field $\dot\phi^a=\varphi^a_2$ as well. However, the new pieces (II) and (III) induce contributions which are one order more suppressed in slow-roll parameters as compared to the first piece (I) leading to negligible corrections for all the trajectories in our two field setup. To see it explicitly, one needs to observe the details of the components of generalized (kind of) metric $A^{\cal A \cal B}$ which have been derived in appendix \ref{expressions}.}

\subsubsection*{Scalar spectral index ($n_S$)}
\begin{figure}[h!]
\centering
\includegraphics[scale=1]{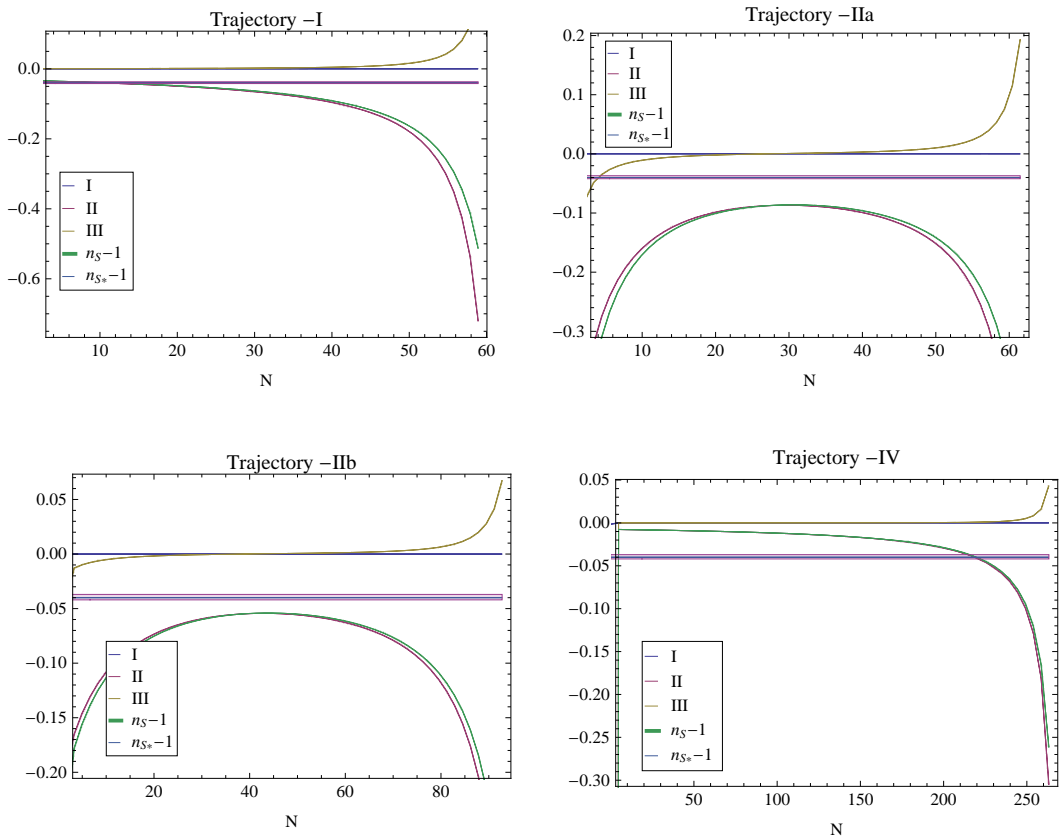}
\caption{Spectral index $n_S-1$ plotted for the four trajectories under consideration. It is observed that the most dominant contribution comes from terms of type-II as mentioned in (\ref{eq:nS-new}). The shadow region shows the allowed window (\ref{obs}) presented by Planck \cite{Planck:2013kta, Ade:2013zuv, Ade:2013ydc, Ade:2013uln}. It is observed that only the trajectories of class I and IV are within the experimental bounds, and class II trajectories (IIa and IIb) are slightly beyond.}
\label{nS}
\end{figure}

The spectral index for scalar perturbation modes of a multi-field inflation driven on a non-flat background can be computed from the relationt
$$n_S -1 = \frac{D\ln P_S}{d\ln k} \simeq \frac{D\ln P_S}{H\, dt} = \frac{D\ln P_S}{d N} ~,~$$
where $\frac{D}{d N}$ is the covariant time derivative along a background trajectory in the field space. Using the general expression (\ref{eq:Ps}) of power spectrum ${\cal P}_S$, we get
\bea
\label{eq:nS}
& & n_S - 1 = - 2 \, \epsilon + 2 \, \frac{A^{\cal A \cal B} \left(\frac{D N_{\cal A}}{d N}\right) \, N_{\cal B}}{A^{\cal A \cal B} \, N_{\cal A} \, N_{\cal B}} +  \frac{ \left(\frac{D A^{\cal A \cal B}}{d N}\right) \, N_{\cal A}\, N_{\cal B}}{A^{\cal A \cal B} \, N_{\cal A} \, N_{\cal B}} ~.~
\eea
For further simplification, we need to utilize the first evolution equation of efolding field derivatives
(\ref{eq:ODEsfieldN}) given as
 $$\frac{D}{dN} { N_{\cal A}}(N) = - { P^{\cal B}_{\, \, \, {\cal A}}}(N) \, { N_{\cal B}}(N),$$
where the explicit expressions for various components of $P^{\cal A}_{\, \, \, \, \cal B}$ are given in (\ref{eq:Pexplicit}). Subsequently, the expression for scalar spectral index simples to
\bea
\label{eq:nS-new}
& & n_S - 1 = - 2 \, \epsilon - 2 \, \frac{A^{\cal A \cal B}  \, N_{\cal A} \, P^{\cal C}_{\, \, \, {\cal B}} \, { N_{\cal C}}}{A^{\cal A \cal B} \, N_{\cal A} \, N_{\cal B}} +  \frac{ \left(\frac{D A^{\cal A \cal B}}{d N}\right) \, N_{\cal A}\, N_{\cal B}}{A^{\cal A \cal B} \, N_{\cal A} \, N_{\cal B}}\\
%& & \nonumber\\
& &   \hskip 2.1cm I \, \, \, \hskip 2.1cm II \, \, \, \hskip 2.1cm III \nonumber
\eea
where we separate out the full expression for $n_S-1$ in three kinds of pieces for numerical investigations. A numerical analysis as reflected in Figure \ref{nS} shows that the first piece (I) is negligible and the most dominant contribution comes from the second piece (II) of Eq.~(\ref{eq:nS-new}). The third piece (III) shows up with some non-trivial values coming from the curvature of the field space generated by $\{\phi^a, \dot\phi^a\}$, however the same does not significantly compete with type II contributions to change the naively expected results. {\it Also, it was observed that for trajectories IIa and IIb, the observed values of scale violation was slightly beyond the experimental bounds.} Besides, larger values indicated in the left most regime of trajectories IIa and IIb  is an outcome of the fact that slow-roll is followed by a fast roll regime which lasts within one or two number of e-foldings as discussed in section~\ref{sec_Setup}.

Although the numerical analysis is done via directly computing the numerical solutions for field derivatives of number of e-foldings, let us elaborate on the expression (\ref{eq:nS-new}) in connection with the literatures. The first two terms of (\ref{eq:nS-new}) are similar to what have been claimed in \cite{Byrnes:2012sc}. The last one is a new type of term which does not appear in \cite{Byrnes:2012sc} because in that case $A^{\cal A \cal B}= A^{ab}_{11}\sim {\cal G}^{ab}$ and metric being a covariantly constant object nullifies the last term. However, for our case the subleading terms are induced which are slow-roll suppressed. Utilizing the explicit expressions of $P^{\cal A}_{\, \, \, {\cal B}}$ (\ref{eq:Pexplicit}), the first two terms in Eq.~(\ref{eq:nS-new}) of spectral index are simplified to the following one
in the slow-roll limit
\bea
\label{eq:nS-new0}
& & n_S - 1 = - 2 \, \epsilon - 2  \frac{\left(\frac{\partial N}{\partial \phi^a}\right)\, \left(\frac{\dot\phi^a \, \dot\phi^d}{H^2} + \frac{1}{3} \, {R^a_{bc}}^d\, \, \frac{\dot\phi^a \, \dot\phi^b}{H^2} - \frac{D^{ad}V}{V}\right) \left(\frac{\partial N}{\partial \phi^d}\right)\,}{{\cal G}^{ab} \, \left(\frac{\partial N}{\partial \phi^a}\right)\, \left(\frac{\partial N}{\partial \phi^b}\right)} ~,~
\eea
which matches with those given in \cite{Sasaki:1995aw, Gong:2002cx}. {Here it is worth to mention that the aforementioned relation is generalized in our approach. It is only the piece of type ${\cal O}^{{\cal A}/{\cal B}} = {\cal O}^{a/b}_1$ of the second part (II) along with the first part (I)  in  our general expression (\ref{eq:nS-new}) which reproduces this result (\ref{eq:nS-new0}) while the terms involving ${\cal O}^{{\cal A}/{\cal B}} = {\cal O}^{a/b}_2$ induce new but subleading contributions. Further, the third piece (\ref{eq:nS-new}) is a new contribution coming from the non-flat metric which are subleading (for the current setup under consideration) but those might be important if the field space is highly non-flat.}

Before getting to the next observable, let us have a very quick cross check for our general formula (\ref{eq:nS-new}) for the simplest single field inflation driven by a scalar field $\phi$ on a flat background. For this case,
Eqs.~(\ref{eq:single1})-(\ref{eq:single3}) along with the
%\bea
%& & N_{A} \equiv \{N_a^1 , N_a^2\} = \{N_\phi, N_{\dot \phi}\}  ~,~\\
%& & P^{\phi}_{\, \, \, \phi}= -\frac{1}{6 H^3} \, \dot \phi \, V_\phi \, , \, \, \, \, \, P^{\dot \phi}_{\, \, \, \phi}= -\frac{V_{\phi \phi}}{ H}+ \frac{1}{6 H^3} \, V_\phi V_\phi  ~,~\nonumber\\
%& & P^{\phi}_{\, \, \,\dot \phi}= \frac{1}{H} -\frac{1}{6 H^3} \, {\dot \phi}^2 \, , \, \, \, \, \, P^{\dot \phi}_{\, \, \, \dot \phi}= -3 \,  + \frac{1}{6 H^3} \, V_\phi \, \dot \phi  ~.~\nonumber
%\eea
slow-roll relations $N_{\dot \phi} \sim \frac{N_\phi}{3 H}$ and $N_\phi = -\frac{H}{\dot\phi}$ immediately imply that
\bea
 \frac{A^{\cal A \cal B}  \, N_{\cal A} \, P^{\cal C}_{\, \, \, {\cal B}} \, { N_{\cal C}}}{A^{\cal A \cal B} \, N_{\cal A} \, N_{\cal B}}  \sim \left(-\frac{{\dot \phi}^2}{H^2} - \frac{V_{\phi \phi}}{V}\right) = 2 \epsilon - \eta_0 ~.~
\eea
Note that here $\frac{D \epsilon}{dN} = 4 \epsilon^2 - 2\, \epsilon \, \eta_0$ where $\eta_0$ is the standard $\eta$ parameter defined as $\eta_0 \equiv \frac{V_{\phi\phi}}{V}$ has been used. After implementing these redefinitions, the scalar spectral index results in
\bea
n_S - 1 \simeq -6 \, \epsilon \, \, + 2 \, \, \eta_0
\eea
which is a well-known standard result for single field case \cite{Cortes:2006ap, Lyth:1998xn}. Note that despite of metric being flat, there are slow-roll suppressed contributions in $A^{\cal A \cal B}$. However, the contribution from the third term in (\ref{eq:nS-new}) would be the second-order slow-roll suppressed.

\subsubsection*{Running of scalar spectral index $n_S$}

\begin{figure}[h!]
\centering
\includegraphics[scale=0.95]{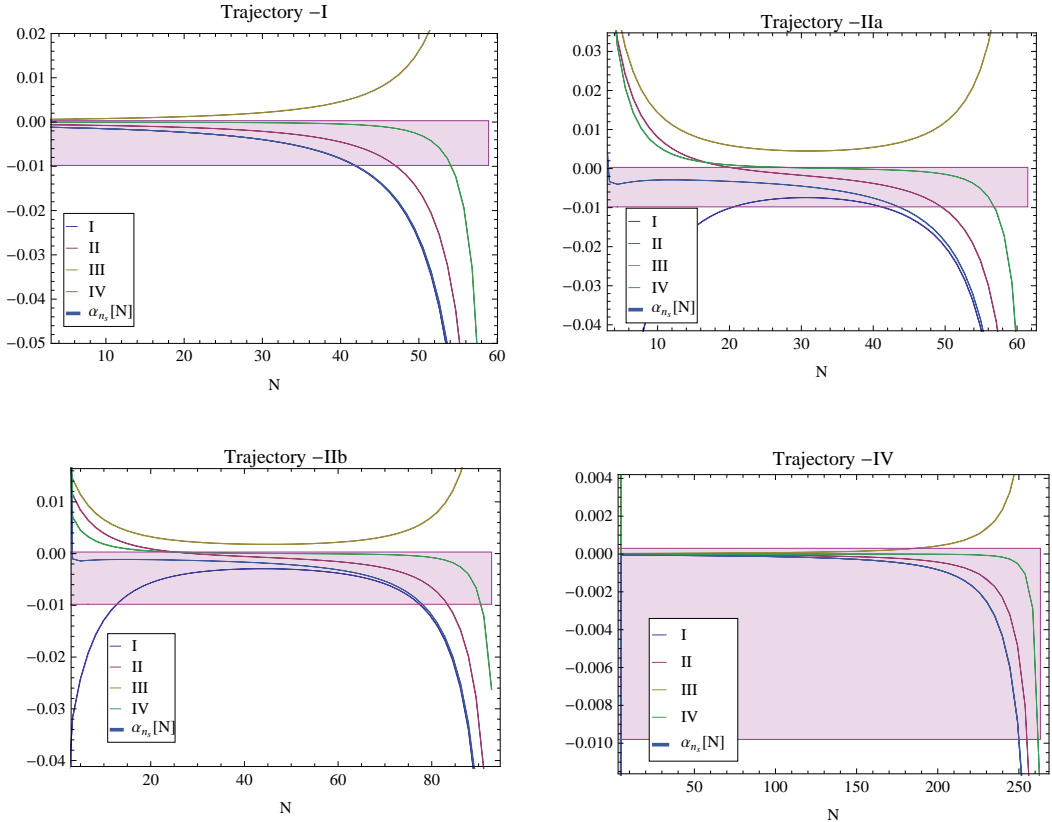}
\caption{Running of spectral index $\alpha_{n_S}$ plotted for the four trajectories under consideration. The shadow region shows the allowed window (\ref{obs}) presented by Planck \cite{Planck:2013kta, Ade:2013zuv, Ade:2013ydc, Ade:2013uln}.}
\label{alphanS}
\end{figure}

Using generic expression for scalar spectral index (\ref{eq:nS-new}), one can easily compute
its running which comes out to be
\bea
\label{eq:alphanS}
& & \hskip-1.5cm \alpha_{n_S} \simeq \frac{D n_S}{d N} = \biggl[- (n_S \, -1 \, + \, 2 \, \epsilon)^2 - 2 \left(\frac{D \, \epsilon}{dN}\right) \biggr]- \,\biggl[\frac{2 \, \left(\frac{D\, P^{\cal B}_{\, \, \, \cal A}}{dN}\right) \, \, \, N^{\cal A} \, N_{\cal B}}{N^{\cal A} \, N_{\cal A}} \biggr] \nonumber\\
	& & \hskip2cm + \biggl[\frac{2}{N^{\cal A} \, N_{\cal A}} \biggl\{ A^{\cal A \cal C} \, P^{D}_{\, \, \, C} \, N_{\cal D} \, \, P^{B}_{\, \, \, A} \, N_{\cal B} + A^{\cal A \cal D} \, P^{B}_{\, \, \, A} \, N_{\cal D} \, \, P^{C}_{\, \, \, B} \, N_{\cal C}\biggr\}\biggr] \\
& & \hskip3cm + \biggl[\frac{1}{N^{\cal A} \, N_{\cal A}} \biggl\{ N_{\cal C}\, \, N_{\cal A} \left(\frac{D^2 A^{\cal A \cal C}}{d N^2} \right) - 2 \, N_{\cal C}\, P^{D}_{\, \, \, A} \, N_{\cal D} \left(\frac{D A^{\cal A \cal C}}{d N}\right) \, \biggr\} \biggr]\nonumber \\
& & \nonumber\\
& & \hskip1cm= (I)  + (II) + (III) + (IV)  ~,~\nonumber
\eea
where each term in big bracket is separated out for numerical comparison given in Figure \ref{alphanS} as under,
A detailed numerical analysis done for the four trajectories under consideration as  plotted in Figure \ref{alphanS} shows that all the pieces I, II, III and IV do have non-trivial contributions, however, their combined effects are well within the experimental bounds.

Before coming to the tensor perturbative modes, let us derive the expression of the running of spectral index $\alpha_{n_s}$ for a single filed inflationary potential. The same would help in understanding the insights of the various components in (\ref{eq:alphanS}). Using the single field analogue of various expressions given in (\ref{eq:single1})-(\ref{eq:single3}), we get the following leading order contributions of various parts of (\ref{eq:alphanS}),
\bea
& & (I) \equiv - 24 \, \epsilon^2 \, + 20 \, \epsilon \, \eta_0 + ....... \, \, , \, \, \,  (II) \equiv -16 \, \epsilon^2 \, + 12 \, \epsilon \, \eta_0 - 2 \, \xi^2 + ...... \, \, , \\
& & \hskip-0.45cm (III) \equiv 16 \, \epsilon^2 \, - 16 \, \epsilon \, \eta_0 + ......... \, \, , \, \, \, (IV) \equiv  -60 \, \epsilon^2 \, \eta_0 + ....... \, \, , \nonumber
\eea
where dots denote the subleading corrections while the standard slow-roll definitions $\eta_0 = \frac{V_{\phi\phi}}{V}$ and $\xi^2 = \frac{V_\phi \, V_{\phi\phi\phi}}{V^2}$ have been used. The sum of these contributions gives
 the well-known single field expression at the leading order as below~\cite{Lyth:1998xn}
\bea
& & \alpha_{n_s} \simeq -24 \, \epsilon^2 + 16 \, \epsilon \, \eta_0 - 2 \, \xi^2 + .......\, \, ,
\eea
which shows that each of the terms except those involving derivative of the field space metric are parts of the overall leading order contributions. {Again it is important to recall that similar to the previous cases, these are only the pieces of type ${\cal O}^{{\cal A}/{\cal B}} = {\cal O}^{a/b}_1$ in our general expression (\ref{eq:alphanS}) which sum up to reproduce this standard result while the terms involving ${\cal O}^{{\cal A}/{\cal B}} = {\cal O}^{a/b}_2$ induces new but subleading contributions with higher order slow-roll suppressed pieces. However the same can not be as clean to observe after expanding out the compact expression leading into too lengthy pieces in terms of  component substituents. Nevertheless in the numerical analysis, these higher order slow-roll effects are automatically included.}

\subsection{Tensor-to-scalar ratio and its scale dependence}
\subsubsection*{Tensor Power spectra ($P_T$)}
The power spectra of the tensor perturbation modes with the
leading order slow-roll correction is given as \cite{Sasaki:1995aw, Lidsey:1995np,Starobinsky:1979ty,Byrnes:2006fr, Gong:2007ha},
\bea
& & {\cal P}_T = 8 \, \left(\frac{H^2}{4 \, \pi^2} \, \left[1-(1+\alpha)\epsilon\right]\right)_{at \, \, \, N = N_*} \, ,
\eea
where $\alpha = 2 - \ln 2 - \gamma \simeq 0.7296$ where $\gamma \simeq 0.5772$ is the Euler-Mascheroni constant \cite{Byrnes:2006fr,Nakamura:1996da, Byrnes:2006vq}.
%\begin{figure}[h!]
%\centering
%\includegraphics[scale=0.85]{Pt.eps}
%\caption{Tensor power spectra $P_T$ plotted for the four trajectories under consideration.}
%\label{Pt}
%\end{figure}
\subsubsection*{Tensorial spectral tilt ($n_T$)}
The spectral tilt for tensor perturbations is defined as \cite{Sasaki:1995aw, Lidsey:1995np,Byrnes:2006fr}
\bea
& & n_T \equiv \frac{D \ln {\cal P}_T}{d\ln k} \simeq \frac{D \ln {\cal P}_T}{d N} \simeq - 2 \, \epsilon - \frac{(1+\alpha) \, \left(\frac{D \epsilon}{dN} \right)}{1-(1+\alpha)\epsilon}  ~.~\,
\eea
% Still need to make sure if this is true for multi-field and non-flat background !
%\begin{figure}[h!]
%\centering
%\includegraphics[scale=0.8]{nT.eps}
%\caption{Tensorial tilt $n_T$ plotted for the four trajectories under consideration.}
%\label{nT}
%\end{figure}
%As observed in Figures \ref{Pt} and \ref{nT},
Here $\frac{D\epsilon}{dN} = 4 \epsilon^2 - 2 \epsilon \, \eta_0$ and so being suppressed by slow-roll parameter $\epsilon$ (which is order $10^{-9}-10^{-7}$ for the four trajectories under consideration), the tensorial tilt is negligibly small for all the trajectories.

%\subsection{Tensor-to-scalar ratio and its scale dependence}
\subsubsection*{Tensor-to-scalar ratio ($r$)}
\label{sec_tensor}
The tensor-to-scalar ratio is one of cosmological parameters which has attracted major attention since long. In general, it is defined as the ratio of power spectra of tensor and scalar perturbation modes and can be written as under \cite{Lidsey:1995np, Byrnes:2006fr,Choudhury:2013iaa}
$$r \equiv \frac{P_T}{P_s}.$$
Using the field derivatives of number of efoldings, we get the following useful relation
\bea
\label{eq:r}
& & r =  8 \, \frac{\left[1-\,(1+\alpha)\epsilon\right]}{N^{\cal A} \, N_{\cal A}}.
\eea
%and the numerical solutions for $N_{\cal A}$ results in the plots given in Figure \ref{rRatio}.
%\begin{figure}[h!]
%\centering
%\includegraphics[scale=1.01]{rRatio.eps}
%\caption{Tensor-to-scalar ratio $r$ plotted for the four trajectories under consideration.}
%\label{rRatio}
%\end{figure}
Also, as it has been elaborated in the appendix \ref{expressions}, the contributions to $r$ as given in (\ref{eq:r}) receive subleading contributions from the $N_b^2$ components of $N^{\cal A} \, N_{\cal A}$. However,
%as observed in Figure \ref{rRatio},
the same still results in a negligibly small value of $r$ for all the trajectories. Neglecting $N_b^2$ component contributions, one gets
\bea
\label{eq:r0}
& & \hskip-2.5cm r  \simeq 8 \, \frac{\left[1- \, (1+\alpha)\epsilon\right]}{N_a^1 \, \left({\cal G}^{ab} - 2 \, \epsilon \, \, {\cal G}^{ab} + 2 \, \alpha \, \frac{{\cal G}^{ac} \epsilon_{cd} N^d_1 \, N^b_1}{{\cal G}^{pq} \, N_p^1\, \, N_q^1}\right) \, N_b^1}  ~~ \, \, \, \, {\rm where}
\eea
$$ \epsilon_{ab} = \epsilon \, {\cal G}_{ab} + \left({\cal G}_{ac} \, {\cal G}_{bd} -\frac{1}{3} \, R_{abcd}\right) \, \frac{\varphi^c_2 \, \varphi^d_2}{H^2} - \frac{V_{;ab}}{3 \, H^2} ~.~$$
\subsubsection*{Running of tensor-to-scalar ratio ($n_r$)}
In \cite{Gong:2007ha}, it was motivated that running of tensor-to-scalar ratio $r$ could be relevant for the detectability through laser interferometer experiments. Based on simple scaling arguments in the power spectra of scalar and tensor perturbations which is
\bea
& & {\cal P}_T \propto k^{n_T} \, \, \, \, {\rm and} \, \, \, \, \, \, {\cal P}_S \propto k^{n_S -1} ~,~
\eea
one gets an overall scale dependence in $r$ given as $r \propto k^{n_T - n_S + 1}$. Therefore, a running in the tensor-to-scalar ratio can be captured as
\bea
\label{eq:r-running}
n_r \equiv \frac{D \ln r}{d \ln k} \simeq \frac{D \ln r}{d N} \equiv 1 - n_S + n_T ~.~
\eea
Further utilizing the expression (\ref{eq:nS-new}), we get the following useful relation
\bea
\label{eq:r-running-new}
n_r \simeq  2 \, \frac{A^{\cal A \cal B}  \, N_{\cal A} \, P^{\cal C}_{\, \, \, {\cal B}} \, { N_{\cal C}}}{A^{\cal A \cal B} \, N_{\cal A} \, N_{\cal B}} -  \frac{ \left(\frac{D A^{\cal A \cal B}}{d N}\right) \, N_{\cal A}\, N_{\cal B}}{A^{\cal A \cal B} \, N_{\cal A} \, N_{\cal B}} ~.~
\eea
%The numerical details for four trajectories are given in Figure \ref{nr}.
%\begin{figure}[h!]
%\centering
%\includegraphics[scale=0.81]{nr.eps}
%\caption{Running of tensor-to-scalar ratio ($n_r$) plotted for the four trajectories.}
%\label{nr}
%\end{figure}
Note that the aforementioned expression (\ref{eq:r-running-new}) consistently reproduces the results of \cite{Gong:2007ha} at the leading order which is
\bea
\label{eq:r-running-new0}
& & n_r = 4 \, \epsilon -2 \, \eta_0 + \frac{2}{3} \, \frac{N_a \, R^a_{bcf} \,{\cal G}^{fd} \, \frac{\partial \phi^b}{\partial N} \,\frac{\partial \phi^c}{\partial N} \, N_d}{{\cal G}^{pq}\, N_p \, N_q} \, \, .
\eea
{As it has been seen throughout, after writing out the quantities in terms of two-fold vectors ${\cal O}_{\cal A} = \{{\cal O}^a_1, {\cal O}^a_2\}$ etc., our expressions generalize the known results at higher order in slow-roll; for example, our tensor-to-scalar ratio given in (\ref{eq:r}) generalizes (\ref{eq:r0}) (given in \cite{Lidsey:1995np, Byrnes:2006fr,Choudhury:2013iaa}) while its running (\ref{eq:r-running-new0}) (given in \cite{Gong:2007ha}) is generalized by our expression (\ref{eq:r-running-new}). Further, the effects of the non-flat background origin can be important in relevant model. The same has not been the case for the present model in which $\epsilon$ parameters are hierarchically smaller than the $\eta$ parameters for all the four trajectories.}

\section{Cosmological observables-II}
\label{sec:cosmo-II}

\subsection{Non-Gaussianity parameters}
\label{sec_fNL}
The signatures of non-Gaussianities are encoded in a set of non-linearity parameters which are commonly denoted as $f_{NL}, \tau_{NL}$ and $g_{NL}$.  These are generically related to the n-point correlators of curvature perturbations; the 2-point correlators simply give rise to a Gaussian shaped power spectrum while the 3-point correlators are related to the bi-spectrum which encodes the non-Gaussianities via the non-linearity parameter $f_{NL}$. Similarly, the 4-point correlators give rise to a tri-spectrum via $\tau_{NL}$ and $g_{NL}$ parameters. Using the $\delta N$-formalism, the non-linearity parameters $f_{NL}$, $\tau_{NL}$ and $g_{NL}$ are defined as,
\bea
\label{eq:fNLgNLtauNLgen}
& &  \hskip-1.0cm f_{NL} = \frac{5}{6} \, \frac{ N^{\cal A} \,  N^{\cal B} \, N_{{\cal A}{\cal B}}}{ (N^{\cal D} \,  N_{\cal D})^2} , \, \, \,  \tau_{NL} = \, \frac{ N^{\cal A} \,  N_{{\cal A}{\cal B}} \, N^{{\cal B}{\cal C}} \, N_{\cal C}}{(N^{\cal D} \,  N_{\cal D})^3} , \, \, \,  g_{NL} = \frac{25}{54} \, \frac{ N^{\cal A} \,  N^{\cal B} \, N^{\cal C} \, N_{{\cal A}{\cal B}{\cal C}}}{ (N^{\cal D} \,  N_{\cal D})^3} \, \, .
\eea
%For the time being, let us proceed with a simple derivation of the generic expressions of running of the non-linearity parameters $f_{NL}, \tau_{NL}$ and $g_{NL}$, and later on we will focus on the numerical approach via computing the field variations of e-fold.
Based on expected hierarchial contributions, we separate out the four contributions of $f_{NL}$ from the generic expression (\ref{eq:fNLgNLtauNLgen}) as below
\bea
\label{eq:fNL0}
& & \hskip-0.4cm f_{NL} = \biggl[\frac{5}{6} \, \frac{ N^{a}_1 \,  N^{b}_1 \, N^{11}_{ab}}{ (N^{\cal D} \,  N_{\cal D})^2}\biggr] +\biggl[\frac{5}{6} \, \frac{ N^{a}_2 \,  N^{b}_1 \, N^{21}_{ab}}{ (N^{\cal D} \,  N_{\cal D})^2}\biggr]+\biggl[\frac{5}{6} \, \frac{ N^{a}_1 \,  N^{b}_2 \, N^{12}_{ab}}{ (N^{\cal D} \,  N_{\cal D})^2}\biggr]+\biggl[\frac{5}{6} \, \frac{ N^{a}_2 \,  N^{b}_2 \, N^{22}_{ab}}{ (N^{\cal D} \,  N_{\cal D})^2}\biggr] \nonumber\\
& & \hskip0.4cm = I + II + III +IV \, \, .
\eea
\begin{figure}[h!]
\centering
\includegraphics[scale=0.81]{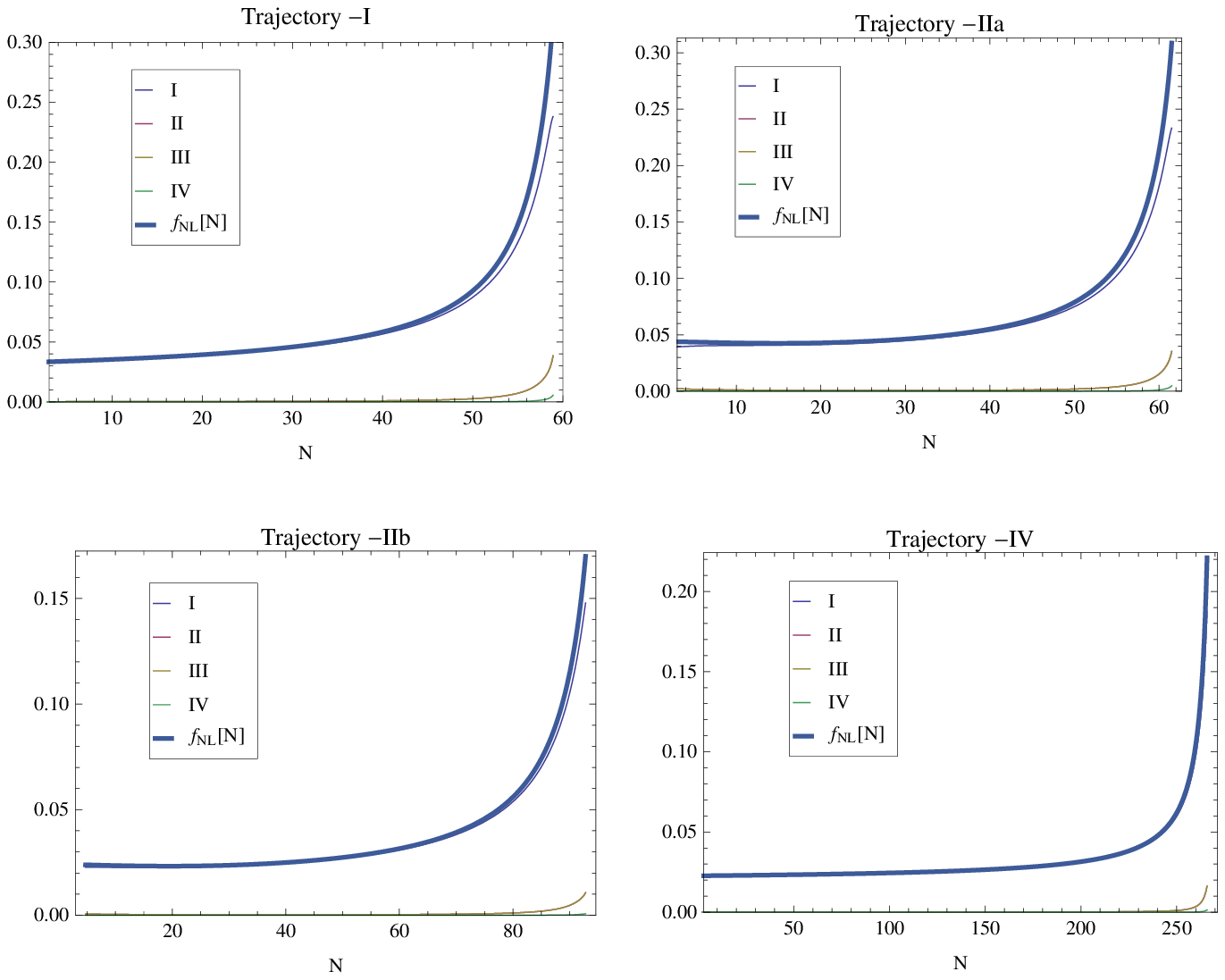}
\caption{Non-linearity parameter $f_{NL}$ plotted for the four trajectories.}
\label{fNL}
\end{figure}
For single field case, using the followings leading order contributions in slow-roll expansion,
\bea
\label{eq:NaAndNab}
& & N_{\cal A}: ~~~~~~~~ N_\phi =- \frac{H}{\dot\phi}, \, \, N_{\dot\phi} \simeq -\frac{N_\phi}{3 \, H},  \\
& & N_{\cal A \cal B}: ~~~~~~~ N_{\phi\phi} \simeq 1-\frac{\eta_0}{2 \epsilon}, N_{\phi\dot\phi} \simeq \frac{1}{3 H} - \frac{\eta_0}{6 H \epsilon} \simeq N_{\dot\phi \phi}, \, N_{\dot\phi \dot\phi} \simeq \frac{1}{6 H^2} - \frac{\eta_0}{18 H^2 \epsilon}\nonumber\\
& & N_{\cal A \cal B \cal C}: ~~~~~N_{\phi \phi \phi} \simeq \frac{-2 \eta _0^2+\xi ^2+2 \eta _0 \epsilon }{2 \sqrt{2} \epsilon ^{3/2}},  N_{\phi \phi \dot\phi} \simeq \frac{-6 \eta _0^2+3 \xi ^2-6 \epsilon^2+9 \eta _0 \epsilon }{18 \sqrt{2} H \epsilon ^{3/2}} \, ,\nonumber\\
& & ~~~~~~~~~~~~~~~N_{\phi \dot\phi \dot\phi} \simeq \frac{-6 \eta _0^2+3 \xi ^2+6 \eta _0 \epsilon }{54 \sqrt{2} H^2 \epsilon ^{3/2}}, \, N_{\dot\phi \dot\phi \dot\phi} \simeq \frac{-6 \eta _0^2+3 \xi ^2-3 \epsilon^2+3 \left(\eta _0-3\right) \epsilon }{162 \sqrt{2} H^3 \epsilon ^{3/2}} \nonumber
\eea
the same results in the following single field expression of $f_{NL}$ parameter
\bea
\frac{6}{5} f_{NL} = 2 \epsilon - \eta_0
\eea
which is a standard result \cite{Byrnes:2006vq}. Note that from Figure \ref{fNL}, it is clear that the first part of expression (\ref{eq:fNL0}) is the most dominant contribution. {The other parts (II-IV) are new contributions and can add up significantly to the overall magnitude towards the end of slow-roll regime, however these new contributions are higher order slow-roll suppressed and negligible for the present setup under consideration.}

\begin{figure}[H]
\centering
\includegraphics[scale=0.90]{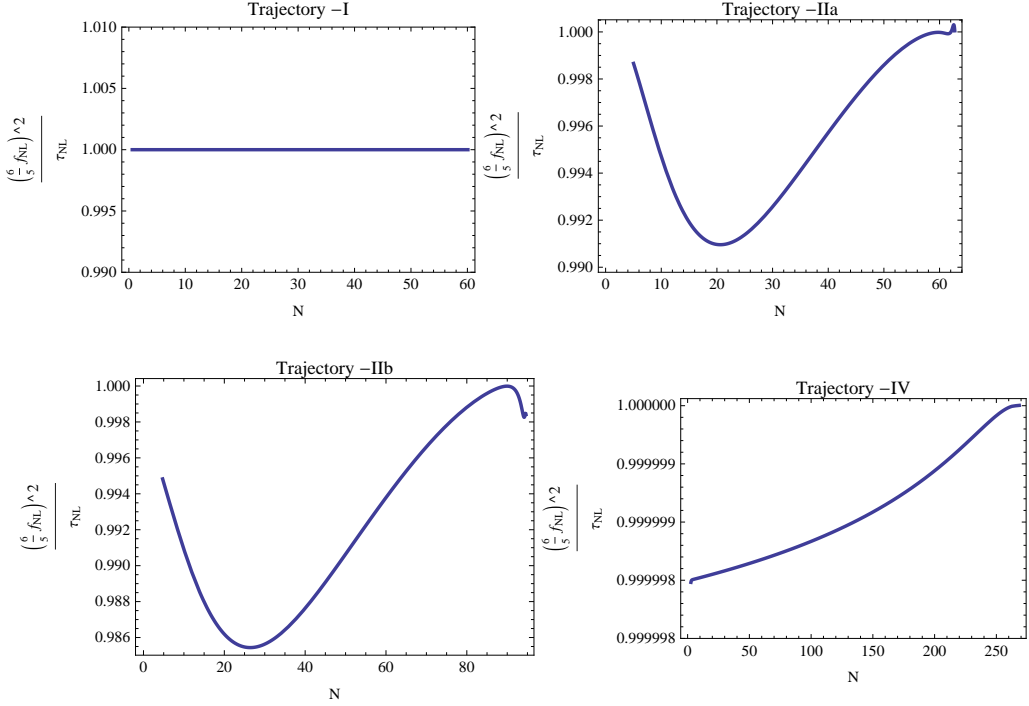}
\caption{Non-linearity ratio parameter $a_{NL}$ plotted for the four trajectories under consideration. The first trajectory being a single field trajectory, there is no deviation from unity. However, at the curving regimes , the other trajectories do have a different values indicating the involvement of multiple fields. }
\label{aNL}
\end{figure}

Similarly, based on expected hierarchial contributions, we separate out the four types of contributions of $\tau_{NL}$, from the definition given in (\ref{eq:fNLgNLtauNLgen}), as below
\bea
& & \hskip-0.8cm \tau_{NL} = \biggl[\frac{ N^{a}_1 \,  N^{11}_{ab} \, N^{bc}_{11} \, N^1_{c}}{(N^{\cal D} \,  N_{\cal D})^3}\biggr] + \biggl[\frac{ N^{a}_2 \,  N^{21}_{ab} \, N^{bc}_{11} \, N^1_{c}}{(N^{\cal D} \,  N_{\cal D})^3} + \frac{ N^{a}_1 \,  N^{12}_{ab} \, N^{bc}_{21} \, N^1_{c}}{(N^{\cal D} \,  N_{\cal D})^3} + \frac{ N^{a}_1 \,  N^{11}_{ab} \, N^{bc}_{12} \, N^2_{c}}{(N^{\cal D} \,  N_{\cal D})^3}\biggr]\nonumber\\
& & \hskip0.1cm + \biggl[\frac{ N^{a}_2 \,  N^{22}_{ab} \, N^{bc}_{21} \, N^1_{c}}{(N^{\cal D} \,  N_{\cal D})^3} +\frac{ N^{a}_1 \,  N^{12}_{ab} \, N^{bc}_{22} \, N^2_{c}}{(N^{\cal D} \,  N_{\cal D})^3} +\frac{ N^{a}_2 \,  N^{21}_{ab} \, N^{bc}_{12} \, N^2_{c}}{(N^{\cal D} \,  N_{\cal D})^3}\biggr] +\biggl[\frac{ N^{a}_2 \,  N^{22}_{ab} \, N^{bc}_{22} \, N^2_{c}}{(N^{\cal D} \,  N_{\cal D})^3}\biggr] \nonumber\\
& & \hskip0.1cm = I + II + III +IV \, \, .
\eea
\begin{figure}[h!]
\centering
\includegraphics[scale=0.90]{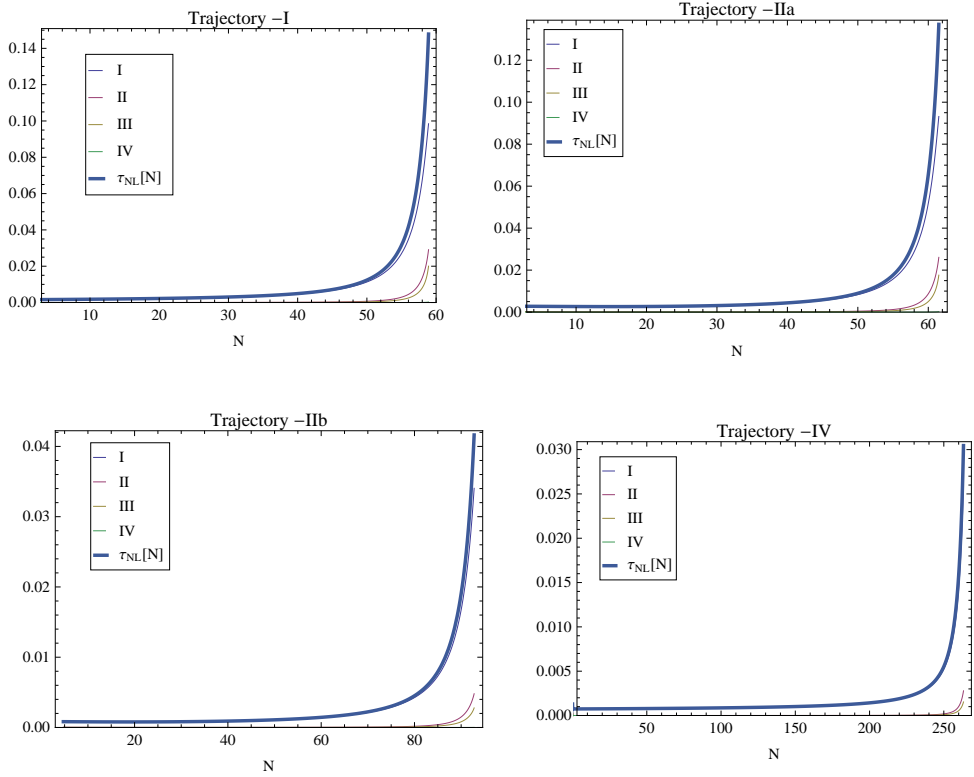}
\caption{Non-linearity parameter $\tau_{NL}$ plotted for the four trajectories.}
\label{tauNL}
\end{figure}
From Figure \ref{tauNL}, it is clear that the first part is the most dominant contribution. As expected, using (\ref{eq:NaAndNab}), one gets the following leading order single field expression \cite{Byrnes:2006vq}
\bea
\tau_{NL} = \left(\frac{6}{5} f_{NL}\right)^2 \simeq \left(2 \epsilon - \eta_0\right)^2.
\eea
Apart from the non-linearity parameters $f_{NL}$ and $\tau_{NL}$, the following relation known as Suyama-Yamaguchi inequality  \cite{Suyama:2007bg}
\bea
\label{eq:aNL}
a_{NL} \equiv \frac{\left(\frac{6}{5} \, f_{NL}\right)^2}{\tau_{NL}} \le 1
\eea
is also of great importance. The equality holds for single field inflationary models. So any deviation of this parameter $a_{NL}$ away from unity automatically indicates a multi-field process happening and then this parameter (along with others) could be a possible discriminator for the known plethora of inflationary models.  The respective numerical details for the four trajectories are given in Figure \ref{aNL}.

\begin{figure}[H]
\centering
\includegraphics[scale=0.95]{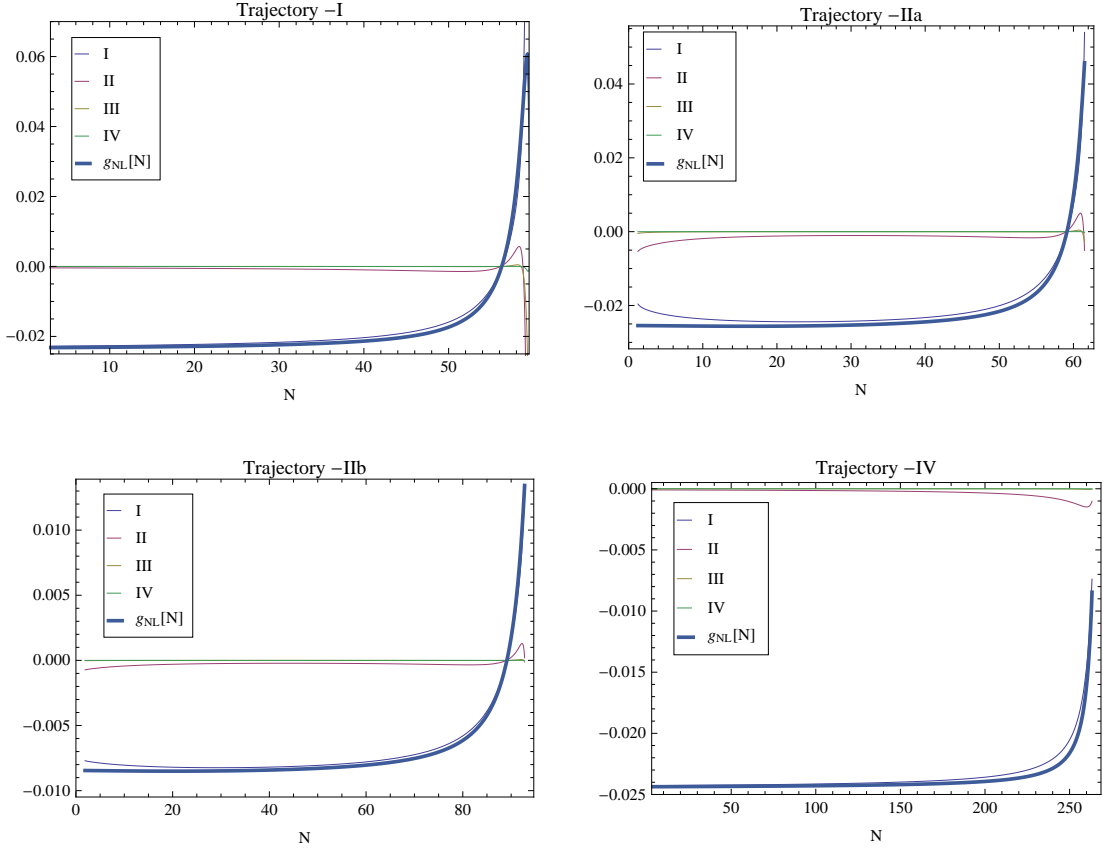}
\caption{Non-linearity parameter $g_{NL}$ plotted for the four trajectories.}
\label{gNL}
\end{figure}

Similarly, according to the expected hierarchial contributions, one can separate out the four contributions of $g_{NL}$ in (\ref{eq:fNLgNLtauNLgen}) also given as below
\bea
\label{eq:gNL0}
& & \hskip-1cm g_{NL} = \biggl[\frac{25}{54} \, \frac{ N^{a}_1  N^{b}_1 N^{c}_1 N^{111}_{abc}}{(N^{\cal D} \,  N_{\cal D})^3}\biggr]\\
& & \hskip0cm +\biggl[\frac{25}{54} \, \frac{\left(N^{a}_2 N^{b}_1 N^{c}_1 N^{211}_{abc}+N^{a}_1 N^{b}_2 N^{c}_1 N^{121}_{abc}+N^{a}_1  N^{b}_1 N^{c}_2 N^{112}_{abc}\right)}{(N^{\cal D} \,  N_{\cal D})^3}\biggr] \nonumber\\
& & \hskip0cm +\biggl[\frac{25}{54} \, \frac{\left(N^{a}_2  N^{b}_2 N^{c}_1 N^{221}_{abc}+N^{a}_1  N^{b}_2 N^{c}_2 N^{122}_{abc}+N^{a}_2  N^{b}_1 N^{c}_2 N^{212}_{abc}\right)}{(N^{\cal D} \,  N_{\cal D})^3}\biggr] \nonumber\\
& & \hskip0cm + \biggl[\frac{25}{54} \, \frac{ N^{a}_2  N^{b}_2 N^{c}_2 N^{222}_{abc}}{(N^{\cal D} \,  N_{\cal D})^3}\biggr] = I + II + III +IV \, \, .\nonumber
\eea
The numerical details for these non-linear parameters as given in Figures \ref{fNL},  \ref{tauNL} and \ref{gNL} indicate that these parameters are negligibly small near the horizon exit and become non-trivial only towards the end of inflation where $\eta$ parameter becomes close to unity. Using (\ref{eq:NaAndNab}), one gets the following standard single field leading order contribution \cite{Byrnes:2006vq}
\bea
\label{eq:gNLnew}
\frac{54}{25} \, g_{NL} \simeq \left(2 \epsilon \, \eta_0 - 2 \eta_0^2 + \xi^2 \right) + .....
\eea
{Thus our expression (\ref{eq:gNL0}) generalizes earlier result of expression (\ref{eq:gNLnew}), the one given in \cite{Byrnes:2006vq}, with the new terms being (II-IV).}

%The same happens because of an exponential hierarchy between $\epsilon$ and $\eta$ parameters which was the reason for realizing large non-Gaussianities in beyond slow-roll regime in \cite{Gao:2013hn}.

\subsection{Running of non-Gaussianity parameters}
\subsubsection*{Running of $f_{NL}$} 
\begin{figure}[h!]
\centering
\includegraphics[scale=0.93]{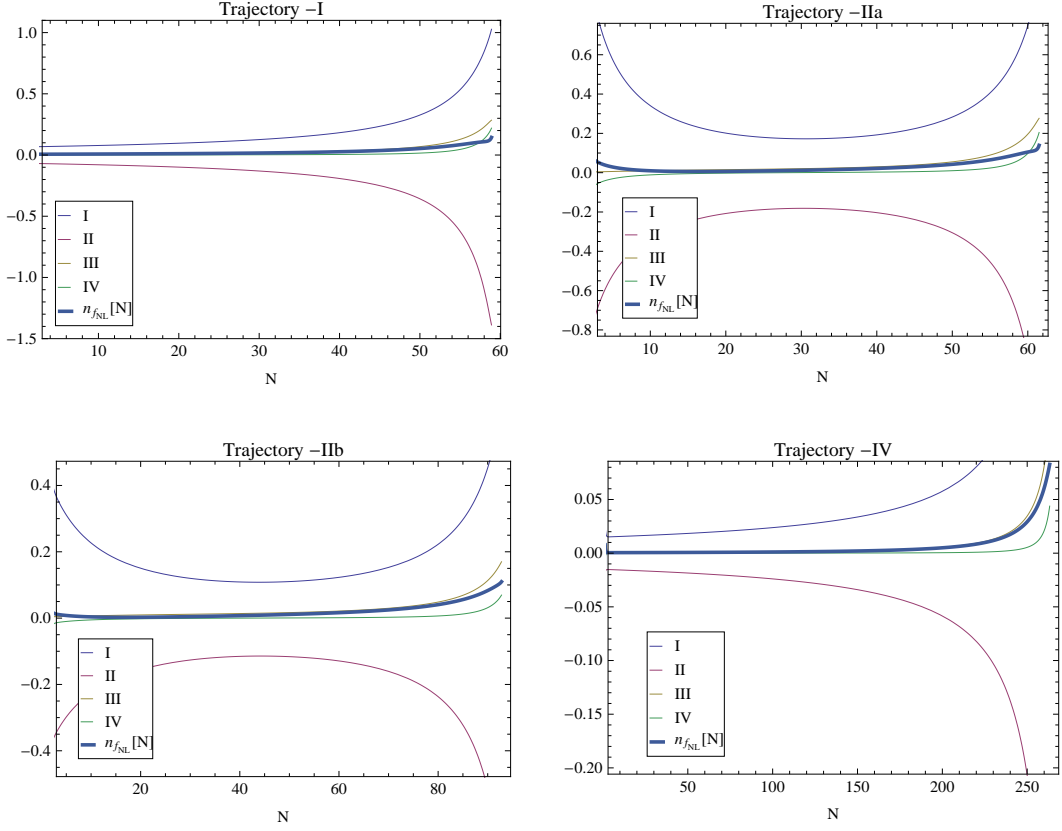}
\caption{Running of $f_{NL}$ plotted for four trajectories under consideration.}
\label{nfNL}
\end{figure}

Using (\ref{eq:fNLgNLtauNLgen}), the running of $f_{NL}$ can be computed as
\bea
\label{eq:nfNL1}
& &\hskip-1.2cm  n_{f_{NL}} \equiv \frac{D \ln f_{NL}}{dk} \sim \frac{D \ln f_{NL}}{dN} \\
& & = -4 \, \, \frac{A^{\cal C \cal D} \, \left(\frac{D N_{\cal C}}{dN}\right) \, N_{\cal D}}{A^{\cal A \cal B} \, N_{\cal A} \, N_{\cal B}} + 2 \, \, \frac{A^{\cal C \cal B} \,\left(\frac{D N_{\cal B}}{dN}\right) \, N^{\cal D} \, N_{\cal C \cal D}}{N_{\cal A \cal B} \, N^{\cal A} \, N^{\cal B}} + \frac{\left(\frac{D N_{\cal C \cal D}}{dN}\right) \, N^{\cal C} \, N^{\cal D}}{N_{\cal A \cal B} \, N^{\cal A} \, N^{\cal B}} \nonumber\\
& & \hskip2cm  -2 \, \, \frac{\left(\frac{D A^{\cal C \cal D}}{dN}\right) \, N_{\cal C} \, N_{\cal D}}{A^{\cal A \cal B} \, N_{\cal A} \, N_{\cal B}} +  \, \, \frac{\left(\frac{D A^{\cal C \cal B}}{dN}\right)\, N_{\cal B} \, N^{\cal D} \, N_{\cal C \cal D}}{N_{\cal A \cal B} \, N^{\cal A} \, N^{\cal B}}   ~.~\nonumber
\eea
Now utilizing the first two evolution equations of (\ref{eq:ODEsfieldN}) for $N_{\cal A}$ and $N_{\cal A \cal B}$ given as follows
\bea
& & \frac{D}{dN} { N_{\cal A}}(N) = - { P^{\cal B}_{\, \, \, {\cal A}}}(N) \, { N_{\cal B}}(N) ~,~ \nonumber\\
& & \frac{D}{dN} { N_{\cal A \cal B}}(N)  = - N_{\cal A \cal C} \, P^{\cal C}_{\, \, \, {\cal B}} - N_{\cal B \cal C} \, P^{\cal C}_{\, \, \, {\cal A}}  - N_{\cal C} \, Q^{\cal C}_{\, \, \, \cal A \, {\cal B}} ~,~ \nonumber
%& & \frac{D}{dN} { N_{\cal A \cal B \cal C}}(N)  = - N_{\cal D} \, Q^{\cal D}_{\, \, \, \cal A \, {\cal B} \, \cal C} - N_{\cal A \, {\cal D}} \, Q^{\cal D}_{\, \,  \, {\cal B} \, \cal C}- N_{\cal B \, {\cal D}} \, Q^{\cal D}_{\, \, \, \cal A  \, \cal C}- N_{\cal C \, {\cal D}} \, Q^{\cal D}_{\, \, \, \cal A \, {\cal B}} \nonumber\\
%& & \, \, \, {\rm yet \, \, to \, \, compute} \, \, ??? \nonumber
\eea
the expression (\ref{eq:nfNL1}) for $n_{f_{NL}}$ is simplified to the one given below
\bea
\label{eq:nfNL2}
& &\hskip-1cm  n_{f_{NL}} =  4 \, \, \frac{P^{\cal B}_{\, \, \, {\cal D}} \, N_{\cal B} \, N^{\cal D}}{N^{\cal D} \, N_{\cal D}} - 2 \, \, \frac{P^{\cal A}_{\, \, \, {\cal C}} \, N^{\cal C} \, N^{\cal B}\, N_{\cal A \cal B}}{N^{\cal C} \, N^{\cal D} \, N_{\cal C \cal D}} - 2 \, \, \frac{P^{\cal D}_{\, \, \, {\cal C}} \, N_{\cal D} \, N^{\cal B}\, N^{\cal C}_{ \, \, \cal B}}{N^{\cal C} \, N^{\cal D} \, N_{\cal C \cal D}} \nonumber
\eea
\bea
& & \hskip1.2cm - \frac{N^{\cal A} \, N^{\cal B} \, Q^{\cal C}_{\, \, \, \, \cal A \cal B} \, N_{\cal C}}{N^{\cal C} \, N^{\cal D} \, N_{\cal C \cal D}} -2 \, \, \frac{\left(\frac{D A^{\cal C \cal D}}{dN}\right) \, N_{\cal C} \, N_{\cal D}}{N^{\cal A} \, N_{\cal A}} +  \, \, \frac{\left(\frac{D A^{\cal C \cal B}}{dN}\right)\, N_{\cal B} \, N^{\cal D} \, N_{\cal C \cal D}}{N_{\cal A \cal B} \, N^{\cal A} \, N^{\cal B}} ~.~
\eea
Further using the expression of scalar spectral index (\ref{eq:nS-new}), it is good to point out that our expression of running of $f_{NL}$ can be written as a generalized version to that of \cite{Leung:2013rza} as below
\bea
\label{eq:nfNL}
& & n_{f_{NL}} = -\biggl[2\, (n_S - 1 + 2 \, \epsilon)\biggr]-2 \, \biggl[ \frac{P^{\cal A}_{\, \, \, {\cal C}} \, N^{\cal C} \, N^{\cal B}\, N_{\cal A \cal B}}{N^{\cal C} \, N^{\cal D} \, N_{\cal C \cal D}} + \, \frac{P^{\cal D}_{\, \, \, {\cal C}} \, N_{\cal D} \, N^{\cal B}\, N^{\cal C}_{ \, \, \cal B}}{N^{\cal C} \, N^{\cal D} \, N_{\cal C \cal D}} \,  \biggr]\nonumber\\
& & \hskip2cm -\biggl[ \frac{N^{\cal A} \, N^{\cal B} \, Q^{\cal C}_{\, \, \, \, \cal A \cal B} \, N_{\cal C}}{N^{\cal C} \, N^{\cal D} \, N_{\cal C \cal D}} \biggr] +  \, \, \biggl[\frac{\left(\frac{D A^{\cal C \cal B}}{dN}\right)\, N_{\cal B} \, N^{\cal D} \, N_{\cal C \cal D}}{N_{\cal A \cal B} \, N^{\cal A} \, N^{\cal B}}\biggr] \, \\
& & \hskip1cm = (I) + (II) + (III) + (IV) ~.~ \nonumber
\eea
The first three terms are the generalized version to those given in \cite{Byrnes:2012sc}. Again the last terms is an entirely new and did not appear in the expression given in \cite{Byrnes:2012sc}, since $A^{ab}_{11}\sim {\cal G}^{ab}$ nullifies the term $\frac{D A^{\cal C \cal D}}{dN}$. The numerical details for four trajectories are given in Figure \ref{nfNL} which indicate that $n_{f_{NL}}$ are non-trivial only towards the end of inflation where $\eta$ parameter becomes close to unity. For the single field inflationary potential $V(\phi)$, using (\ref{eq:single1}-\ref{eq:single3}) and (\ref{eq:NaAndNab}) one gets the following leading order contributions,
\bea
(I) \simeq 4(2 \epsilon - \eta_0), ~~ (II) \simeq \frac{-16 \, \epsilon^2 +16 \, \epsilon ~\eta_0 - 4 \, \eta_0^2}{2 \epsilon -\eta_0}, ~~ (III) \simeq \frac{8~\epsilon^2 - 6 ~ \epsilon \, \eta_0 + \xi^2}{2 \epsilon -\eta_0}
\eea
where $(IV)$ is one order more suppressed in slow-roll parameters.  The first three contributions sum to the following well known leading order expression \cite{Byrnes:2009pe}
\bea
n_{f_{NL}} \simeq \frac{8 \epsilon^2 - 6 \epsilon \, \eta_0 + \xi^2}{2 \epsilon - \eta_0} + ..........
\eea
which is standard result. Here, a factor of $(2 \, \epsilon - \eta_0)$ appears from the relation $N_{\cal A \cal B}\, N^{\cal A} \, N^{\cal B} \simeq \frac{2 \epsilon - \eta_0}{ 4 \, \epsilon^2}$ in the denominator of (\ref{eq:nfNL}). {It is worth to mention that our expression (\ref{eq:nfNL}) of running of $f_{NL}$ generalizes the one given in \cite{Leung:2013rza, Byrnes:2009pe} on the same lines of new terms with higher order slow-roll suppression as explained for the previous cases.}

%\subsection{$\tau_{NL}$ and it's running $n_{\tau_{NL}}$}
\subsubsection*{Running of $\tau_{NL}$} 
\begin{figure}[h!]
\centering
\includegraphics[scale=0.95]{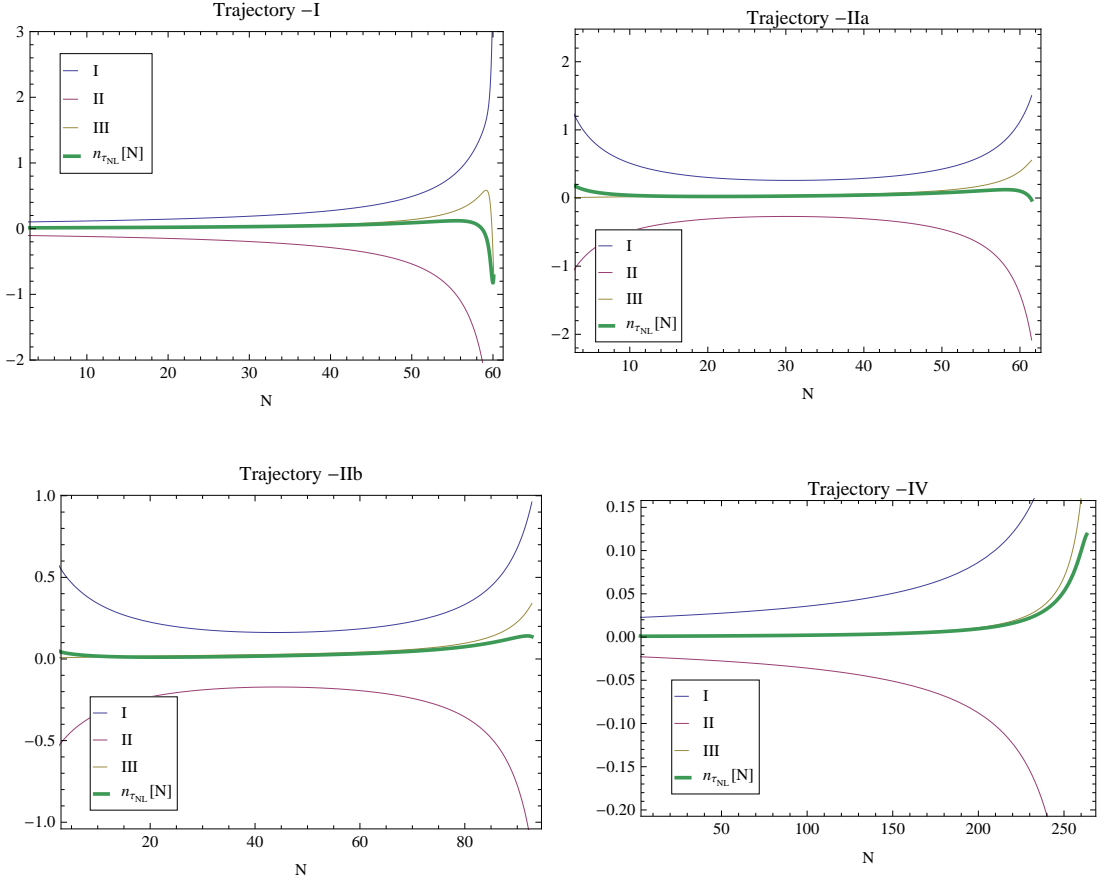}
\caption{Running of $\tau_{NL}$ plotted for four trajectories under consideration.}
\label{ntauNL}
\end{figure}

Using (\ref{eq:fNLgNLtauNLgen}), the running of $\tau_{NL}$ can be represented as
\bea
\label{eq:ntauNL}
& & \hskip-0.7cm n_{\tau_{NL}} \equiv \frac{D \ln \tau_{NL}}{dk} \sim \frac{D \ln \tau_{NL}}{dN} \\
& & = \biggl[6 \, \, \frac{P^{\cal A}_{\, \, \, \, D} \, N_{\cal A} \, N^{\cal D}}{N^{\cal D} \, N_{\cal D}} -3 \, \, \frac{\left(\frac{D A^{\cal C \cal D}}{dN}\right) \, N_{\cal C} \, N_{\cal D}}{N^{\cal A} \, N_{\cal A}} \biggr] -\biggl[2 \, \, \frac{N^{\cal A}\, Q^{\cal D}_{\, \, \, \, \, \cal A \cal B}\, N_{\cal D} \, N^{\cal B \cal C} \, N_{\cal C}}{N^{\cal A}\, N_{\cal A \cal B} \, N^{\cal B \cal C} \, N_{\cal C}}\biggr] \nonumber\\
& & - \biggl[\frac{2 \, N^{\cal A} \, N_{\cal A \cal B} \, N^{\cal B \, \cal C} \, P^{\cal D}_{\, \, \, C} \, N_{\cal D}}{N^{\cal A}\, N_{\cal A \, \cal B}\, N^{\cal B \, \cal C} \, N_{\cal C}} + \frac{2 \, N^{\cal A} \, N_{\cal A \cal D} \, N^{\cal B \, \cal C} \, P^{\cal D}_{\, \, \, B} \, N_{\cal C}}{N^{\cal A}\, N_{\cal A \, \cal B}\, N^{\cal B \, \cal C} \, N_{\cal C}} + \frac{2 \, N^{\cal A} \, N_{\cal B \cal D} \, N^{\cal B \, \cal C} \, P^{\cal D}_{\, \, \, A} \, N_{\cal D}}{N^{\cal A}\, N_{\cal A \, \cal B}\, N^{\cal B \, \cal C} \, N_{\cal C}}\biggr]  \nonumber\\
& & -\bigg[ \frac{2 \, \left(\frac{DA^{\cal A \cal D}}{dN}\right)\, N_{\cal D} \, N_{\cal A \cal B} \, N^{\cal B \cal C} \, N_{\cal C}}{N^{\cal A}\, N_{\cal A \, \cal B}\, N^{\cal B \, \cal C} \, N_{\cal C}}- \frac{2 \, \left(\frac{D(A^{\cal B \cal E} \, A^{\cal C \cal F})}{dN}\right)\, N^{\cal A} \, N_{\cal A \cal B} \, N_{\cal E \cal F} \, N_{\cal C}}{N^{\cal A}\, N_{\cal A \, \cal B}\, N^{\cal B \, \cal C} \, N_{\cal C}}\biggr] \nonumber\\
& & = I + II + III + IV~.~\nonumber
\eea
Again, using the expression of scalar spectral index (\ref{eq:nS-new}), the first bracket terms in (\ref{eq:ntauNL}) reduces to $-3\, (n_S - 1 + 2 \, \epsilon)$, and thus our expression of running of $\tau_{NL}$ receives an analogous form to that of \cite{Leung:2013rza}. The numerical details for four trajectories are given in Figure \ref{ntauNL} which indicate that $n_{\tau_{NL}}$ are non-trivial only towards the end of inflation where $\eta$ parameter becomes close to unity. For the single field inflationary potential $V(\phi)$, using (\ref{eq:single1}-\ref{eq:single3}) and (\ref{eq:NaAndNab}) one gets the following leading order contributions \cite{Leung:2013rza},
\bea
n_{\tau_{NL}} = {2} \, n_{f_{NL}}.
\eea

%\subsection{$g_{NL}$ and it's running $n_{g_{NL}}$}
\subsubsection*{Running of $g_{NL}$}
\begin{figure}[h!]
\centering
\includegraphics[scale=0.95]{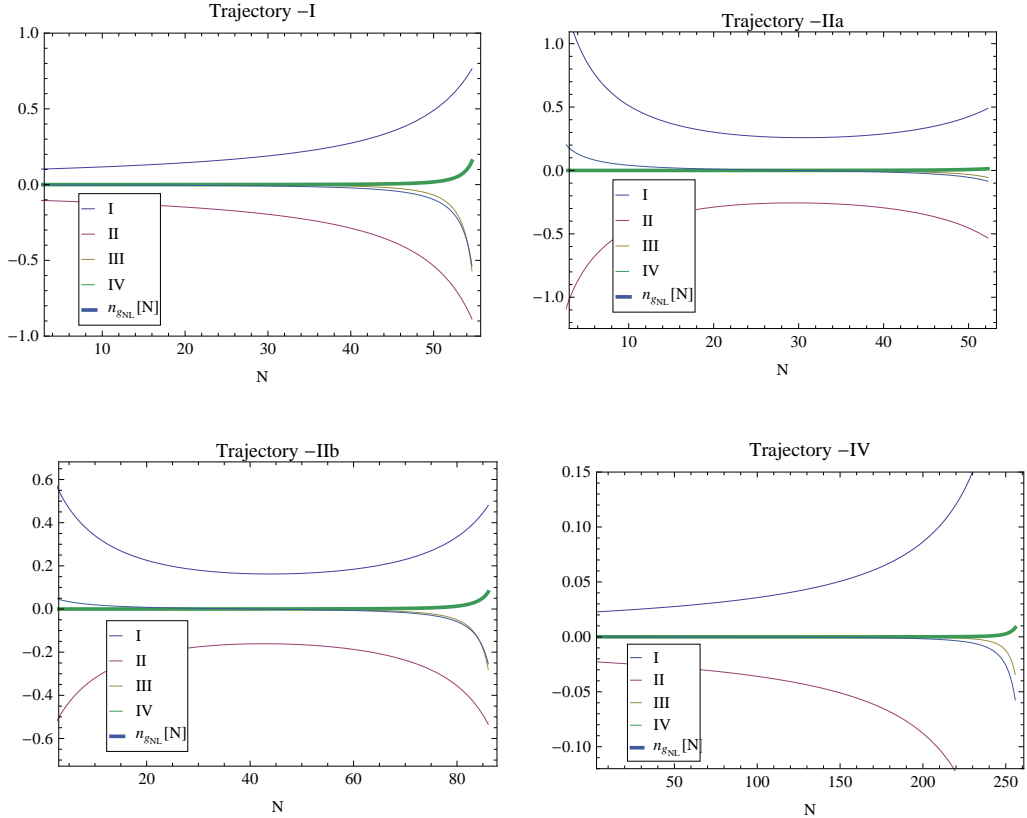}
\caption{Running of $g_{NL}$ plotted for four trajectories under consideration.}
\label{ngNL}
\end{figure}

Using (\ref{eq:fNLgNLtauNLgen}), the running of $g_{NL}$ can be represented as
\bea
& & \hskip-2.3cm n_{g_{NL}} \equiv \frac{D \ln g_{NL}}{dk} \sim \frac{D \ln g_{NL}}{dN} = 6 \, \, \frac{P^{\cal A}_{\, \, \, \, D} \, N_{\cal A} \, N^{\cal D}}{N^{\cal D} \, N_{\cal D}}-3 \, \, \frac{\left(\frac{D A^{\cal C \cal D}}{dN}\right) \, N_{\cal C} \, N_{\cal D}}{N^{\cal A} \, N_{\cal A}} \\
& & \hskip3cm  - \frac{3 \, P^{\cal A}_{\, \, \, \cal D} \, N^{\cal D} \, N^{\cal B} \, N^{\cal C} \, N_{\cal A \cal B \cal C}}{N_{\cal A \, \cal B \, \cal C}\, N^{\cal A} \, N^{\cal B} \, N^{\cal C}} +  \frac{\left(\frac{D N_{\cal A \cal B \cal C}}{dN}\right) \, N^{\cal A} \, N^{\cal B} \, N^{ \cal C}}{N_{\cal A \, \cal B \, \cal C}\, N^{\cal A} \, N^{\cal B} \, N^{\cal C}} ~.~ \nonumber
\eea
To simplify the aforementioned running of $g_{NL}$, we use equation ~(\ref{eq:ODEsfieldN}) to get the following

\bea
\label{eq:ngNL}
& & \hskip-0.75cm n_{g_{NL}}\simeq \biggl[-3\, (n_S - 1 + 2 \, \epsilon) \biggr] -  \biggl[\frac{ \left(Q^{\cal D}_{\, \,  \, {\cal A} \, \cal B}\,N_{\cal D \, {\cal C}} + Q^{\cal D}_{\, \,  \, {\cal B} \, \cal C}\,N_{\cal D \, {\cal A}} + Q^{\cal D}_{\, \,  \, {\cal C} \, \cal A} \,N_{\cal D \, {\cal B}}\right)\, N^{\cal A} \, N^{\cal B} \, N^{\cal C}}{\, N^{\cal A \, \cal B \, \cal C}\, N_{\cal A} \, N_{\cal B} \, N_{\cal C}} \,\biggr] \nonumber\\
& & \hskip0.25cm - \biggl[\frac{3 \, P^{\cal A}_{\, \, \, \cal D} \, N^{\cal D} \, N^{\cal B} \, N^{\cal C} \, N_{\cal A \cal B \cal C} +\left(N_{\cal A \cal B \cal D} \, P^{\cal D}_{\, \, \, \cal C} + N_{\cal A \cal D \cal C} \, P^{\cal D}_{\, \, \, \cal B} + N_{\cal D \cal B \cal D} \, P^{\cal D}_{\, \, \, \cal A}\right)\, N^{\cal A} \, N^{\cal B} \, N^{\cal C}}{N^{\cal A \, \cal B \, \cal C}\, N_{\cal A} \, N_{\cal B} \, N_{\cal C}} \,\biggr] \nonumber\\
& &\hskip0.3cm  - \biggl[\frac{\, \, Q^{\cal D}_{\, \,  \, {\cal A} \, \cal B \, \cal C} \,N_{\cal D} \, N^{\cal A} \, N^{\cal B} \, N^{\cal C}}{N^{\cal A \, \cal B \, \cal C}\, N_{\cal A} \, N_{\cal B} \, N_{\cal C}} \,\biggr]  = I + II + III + IV  ~,~
\eea
%where the dots correspond to the subleasing terms which involve the covariant derivative of the `generalized' metric $\frac{D A^{\cal A \, \cal B}}{d N}$.
where we have neglected the terms with derivatives of $A^{\cal A \cal B}$ as those are found to be negligible in all the previous analysis. The numerical details for four trajectories are given in Figure \ref{ngNL} which indicate that $n_{g_{NL}}$ are non-trivial only in the regions where $\eta$ parameter becomes close to unity. %For the single field inflationary potential $V(\phi)$, one gets the fullowing realigns at leading order,
%\bea
%n_{g_{NL}} \simeq -(2 \, \epsilon - \eta_0) \, n_{\tau_{NL}} + .......... \simeq -2 \, (2 \, \epsilon - \eta_0) \, n_{f_{NL}}
%\eea

\section{Conclusions}
\label{sec_Conclusions and Discussions}

In this article, we presented generalized analytic expressions for various cosmological observables in the context of a multi-field inflation driven on a non-flat field space. A closer investigation has been made regarding the new/generalized contributions to various cosmological observables coming from the non-trivial field space metric, which appears in the standard kinetic term of the scalar field Lagrangian. Subsequently,  in order to connect our findings with the known results, we recovered the standard results as limiting cases from the analytic expressions we derived.

The basic idea has been to rewrite all the cosmological variables in terms of field derivatives of number of e-foldings $N$ and thereafter to solve the differential equation governing the evolution by utilizing the so-called `backward formalism'. For this purpose, we translated the whole problem in solving for the evolution of field-derivatives of $N$ in form of a set of coupled order-one differential equations for vector $N_{\cal A}$, 2-tensor $N_{\cal A \cal B}$ and 3-tensor $N_{\cal A \cal B \cal C}$ quantities. Following the strategy of Yokoyama et al \cite{Yokoyama:2008by}, each of the index ${\cal A}$ counts as $2 \, n$, where $n$ is the number of scalar fields taking part in the inflationary process. This happens because each second-order differential equations for $n$-inflatons has been equivalently written as the first-order differential equations (\ref{EOM}) for $2 \, n$ number of fields. The same implies that the evolution equations for $N_{\cal A}$ results into  $2 \, n$ differential equations while those of $N_{\cal A \cal B}$ and $N_{\cal A \cal B \cal C}$ result in ${4\, n^2}$ and $8 \, n^3$ order-one differential equations, respectively. This is obvious that the numerical analysis gets difficult for large number of scalar fields involved, however, we exemplified the analytic results for a two-field inflationary model, and hence the analysis still remains well under controlled as well as efficient for solving 84 order-one (but coupled) differential equations.

The analytic expressions of various cosmological observables have been utilized for a detailed numerical analysis in a two field inflationary model realized in the context of large volume scenarios. In this model, the inflationary process is driven by a so-called Wilson divisor volume modulus and its respective $C_4$ axion appearing in the chiral coordinate. The same results in a `roulette' type inflation in which depending on the initial conditions, various inflationary trajectories can generate sufficient number of e-foldings as well as significant curving during the inflationary dynamics. Apart from a consistent realization of CMB results, we have also studied the scale dependence of non-Gaussianity observables which could be interesting from the point of view of upcoming experiments.

The analytic expressions for various cosmological observables derived in this article involve the quantities/intermediate ingredients in the form of ${\cal O}^{\cal A} \equiv \{ {\cal O}^a_1, {\cal O}^a_2\}$. Unlike the usual approach, it includes not only the derivative with respect to the field ${\cal O}^a_1$ but also the derivatives with respect to the time derivatives of the field ${\cal O}^a_2$. This method subsequently induce new terms to generalize the previously known expressions of the respective observables with subleading higher order slow-roll corrections. Moreover, the expressions are derived for any generic multi-field inflationary potential with non-flat background and thus could be applicable and useful for generic models.

\subsubsection*{Acknowledgments}

We gratefully acknowledge the enlightening discussions with Anupam Mazumdar and for being a part of the project in the initial stage. We would like to thank Joseph Elliston, Jinn-Ouk Gong and Jonathan White for very useful conversations. TL and XG was supported in part by the Natural Science Foundation of China under grant numbers 10821504, 11075194, 11135003, and 11275246, and by the National Basic Research Program of China (973 Program) under grant number 2010CB833000. PS was supported by the Compagnia di San Paolo contract ``Modern Application of String Theory'' (MAST) TO-Call3-2012-0088.

%%%%%%%%%%%%%%%%%%%%%%%%%%%%%%%%%%%%%%%%%%%%%%%
%%%%%%%%%%%%%%%%%%%%%%%%%%%%%%%%%%%%%%%%%%%%%%%
%\newpage
\begin{appendix}
\section{Details about various components of $A^{\cal A \cal B}$}
\label{expressions}
\begin{figure}[H]
\centering
\includegraphics[scale=0.8]{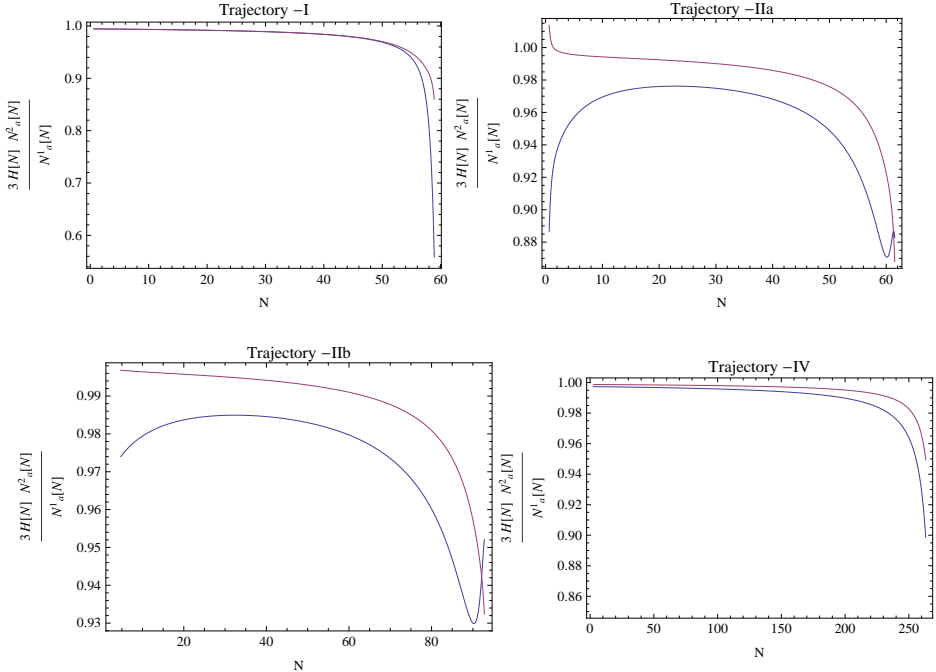}
\caption{Ratio of the two components of $N_a^1$ and $N_a^2$ plotted for the four trajectories. These plots show that in the regime of $\epsilon \ll1$ and $\eta\ll1$, the relation `` $3 \, H \, N_a^2 \sim N_a^1$ " is justified to a reasonably good extent.}
\label{N2vsN1}
\end{figure}

Th role of two tensor $A^{\cal A \cal B}$ is equivalent to a metric in the configuration space generated with the fields $\varphi^a_1$ and $\varphi^a_2$. The same can be generically defined through the following
two-point correlator of field fluctuations $\delta \varphi^{\cal A}$
\bea
\label{AABgen}
& &  \left<\delta \varphi^{\cal A}_* \, \delta \varphi^{\cal B}_*\right> = { A^{{\cal A}{\cal B}}}\, \left(\frac{H_*}{2\pi}\right)^2 .
\eea
In general, $ {A^{{\cal A}{\cal B}}}$ depends on the non-flat background metric as well as on the slow-roll parameters. Up to a good approximation, the two point correlator of $\varphi^a_1$ fluctuations are given as \cite{Nakamura:1996da}
\bea
\label{Aab11}
& & \left<\delta \varphi^{a}_{1*} \, \delta \varphi^{b}_{1*}\right> =  \left(\frac{H_*}{2\pi}\right)^2 \, \left[{\cal G}^{ab} - 2 \, \epsilon \, \, {\cal G}^{ab} + 2 \, \alpha \, \frac{{\cal G}^{ac} \epsilon_{cd} N^d_1 \, N^b_1}{{\cal G}^{pq} \, N_p^1\, \, N_q^1}\right] ~.~
\eea
In the above expression, $\alpha = 2 - \ln 2 - \gamma \simeq 0.7296$ where $\gamma \simeq 0.5772$ is the Euler-Mascheroni constant \cite{Nakamura:1996da, Byrnes:2006vq, Byrnes:2006fr}, and $\epsilon_{ab}$ is defined as
\bea
& & \epsilon_{ab} = \epsilon \, {\cal G}_{ab} + \left({\cal G}_{ac} \, {\cal G}_{bd} -\frac{1}{3} \, R_{abcd}\right) \, \frac{\varphi^c_2 \, \varphi^d_2}{H^2} - \frac{V_{;ab}}{3 \, H^2} ~.~
\eea
Now comparing Eqs.~(\ref{AABgen}) and (\ref{Aab11}), we simply get the component $A^{ab}_{11}$. For getting the other components of $A^{\cal A \cal B}$, let us consider the following form of the Friedmann field equation (\ref{EOM})
\bea
\label{fieldevolution2}
& & \frac{D \, \varphi^a_2}{d \, t} + 3 \, H \, \varphi^a_2 + V^a = 0 ~.~
\eea
The aforementioned evolution equation (\ref{fieldevolution2}) along with the following relation
$$\left(\delta \frac{D}{d t} - \frac{D}{d t} \delta \right)\varphi^a_2 = \left[R^a_{\, \, \,cbd} \, \varphi^c_2 \, \varphi^d_2 \right] \, \delta\varphi^a_1$$
and the slow-roll simplifications, result in the fluctuations of $\delta\varphi^a_2$ to be of the form\footnote{The relation (\ref{deltaphia2}) differs to the analogous expression given in  \cite{Yokoyama:2007dw}, and the difference is due to definition of their $\varphi^a_2 = \frac{d \phi^a}{d N}$ which for our case it is  $\varphi^a_2 = \frac{d \phi^a}{d t}$, and the appearance of curvature corrections.}
\bea
\label{deltaphia2}
& & \delta\varphi^a_2 \simeq \left( \frac{V^a \, V_b}{18 \, H^3} - \frac{V^a_{\, \, \, \,\, ;b}}{3 \, H} + \frac{1}{3 \, H} \, R^a_{\, \, \, cdb} \, \varphi^c_2 \varphi^d_2\right) \, \delta\varphi^b_1 \equiv  \Delta^a_b\,
\delta\varphi^b_1 ~.~
\eea
By using relations (\ref{deltaphia2}) along with  (\ref{AABgen}) and (\ref{Aab11}),  all the components of $A^{\cal A \cal B}$ can be immediately picked up as follows
\bea
\label{Aab}
\label{FullAAB}
& & {A^{ab}_{11}}={\cal G}^{ab} - 2 \, \epsilon \, \, {\cal G}^{ab} + 2 \, \alpha \, \frac{{\cal G}^{ac} \epsilon_{cd} N^d_1 \, N^b_1}{{\cal G}^{pq} \, N_p^1\, \, N_q^1}~;\nonumber\\
& &  {A^{ab}_{12}} = \Delta^a_c \, A^{c b}_{11}= \left({ A^{ab}_{21}}\right)^T  \, \, \, \, \, {\rm and} \, \, \, \, \,  {A^{ab}_{22}} =\Delta^a_c \, \Delta^b_d \, A^{c d}_{11}~.
\eea
Note that, the leading order slow-roll correction to ${A^{ab}_{11}}$ are also consistent with those of \cite{Byrnes:2006vq,Byrnes:2006fr}, for example, with a diagonal field space metric ${\cal G}_{ab}$, the off-diagonal contributions to $A^{ab}_{11}$ appears only with non-standard corrections with coefficient $\alpha$. Also, in slow-roll regime the relations $N_a^2 \sim \frac{N_a^1}{3 \, H}$ holds \cite{Lee:2005bb},  and the same is justified by the plots in Figure \ref{N2vsN1}.
Now utilizing the various components of (\ref{FullAAB}) in $N^{\cal A} = A^{\cal A \cal B} N_{\cal B}$, we get useful relations
\bea
& & N^{a}_1 = A^{ab}_{11} \, N_{b}^1 + A^{ab}_{12} \, N_{b}^2 \simeq \left( A^{ab}_{11}  + \frac{A^{ab}_{12}}{3 \, H}\right) \, N_{b}^1 ~,~ \nonumber\\
& & N^{a}_2 = A^{ab}_{21} \, N_{b}^1 + A^{ab}_{22} \, N_{b}^2 \simeq \left( A^{ab}_{21}  + \frac{A^{ab}_{22}}{3 \, H}\right) \, N_{b}^1 ~.~ \nonumber
\eea
Using the aforementioned relation, one can observe that $A^{ab}_{12}$  and $A^{ab}_{21}$ are suppressed by slow-roll parameters as compared to $A^{ab}_{11}$ while $A^{ab}_{22}$ is suppressed by two orders of slow-roll parameters as compared to $A^{ab}_{11}$.

\section{Single field components of ${{\cal Q}_{(4)}}^{\cal A}_{\, \, \, \, {\cal B}\, {\cal C} \, {\cal D}}$}
\label{sec_QABCD}
The sixteen components of ${{\cal Q}_{(4)}}^{\cal A}_{\, \, \, \, {\cal B}\, {\cal C} \, {\cal D}}$ for single filed potential with flat background
are
\bea
\label{eq:single4}
& & {\cal Q}^{\phi}_{\, \, \, \phi \phi \phi}=  -\frac{5 \dot{\phi } V_{\phi }^3}{72 H^7}+\frac{\dot{\phi } V_{\phi } V_{\phi \phi }}{4 H^5}-\frac{\dot{\phi } V_{\phi \phi \phi }}{6 H^3} , \, \, \, \, \, {\cal Q}^{\dot\phi}_{\, \, \, \phi \dot\phi \phi}= \frac{5 \dot{\phi } V_{\phi }^3}{72 H^7}-\frac{\dot{\phi } V_{\phi } V_{\phi \phi }}{4 H^5}+\frac{\dot{\phi } V_{\phi \phi \phi }}{6 H^3}   ~,~\nonumber\\
& & {\cal Q}^{\phi}_{\, \, \,\dot\phi \phi \phi}= -\frac{5 \dot{\phi }^2 V_{\phi }^2}{72 H^7}+\frac{\dot{\phi }^2 V_{\phi \phi }}{12 H^5}+\frac{V_{\phi }^2}{12
   H^5}-\frac{V_{\phi \phi }}{6 H^3} \, , \, \, \, \, \, {\cal Q}^{\phi}_{\, \, \,\dot\phi \dot\phi \phi}=  \frac{\dot{\phi } V_{\phi }}{4 H^5}-\frac{5 \dot{\phi }^3 V_{\phi }}{72 H^7}~,~\nonumber\\
& & {\cal Q}^{\dot\phi}_{\, \, \, \phi \phi \phi}=  \frac{5 V_{\phi }^4}{72 H^7}-\frac{V_{\phi }^2 V_{\phi \phi }}{2 H^5}+\frac{2 V_{\phi } V_{\phi \phi \phi }}{3
   H^3}+\frac{V_{\phi \phi }^2}{2 H^3}-\frac{V_{\phi \phi \phi \phi }}{H}\, , \, \, \, \, \,  {\cal Q}^{\phi}_{\, \, \, \phi \dot\phi \dot\phi}= \frac{\dot{\phi } V_{\phi }}{4 H^5}-\frac{5 \dot{\phi }^3 V_{\phi }}{72 H^7} ~,~\\
& & {\cal Q}^{\phi}_{\, \, \, \phi \dot\phi \phi}=  -\frac{5 \dot{\phi }^2 V_{\phi }^2}{72 H^7}+\frac{\dot{\phi }^2 V_{\phi \phi}}{12 H^5}+\frac{V_{\phi }^2}{12 H^5}-\frac{V_{\phi \phi }}{6 H^3} \, , \, \, \, \, \, {\cal Q}^{\dot\phi}_{\, \, \, \dot\phi \dot\phi \dot\phi}= \frac{5 \dot{\phi }^3 V_{\phi }}{72 H^7}-\frac{\dot{\phi} V_{\phi }}{4 H^5}  ~ ,~\nonumber\\
& & {\cal Q}^{\phi}_{\, \, \, \phi \phi \dot\phi}=  -\frac{5 \dot{\phi }^2 V_{\phi }^2}{72 H^7}+\frac{\dot{\phi }^2 V_{\phi \phi }}{12 H^5}+\frac{V_{\phi }^2}{12 H^5}-\frac{V_{\phi
   \phi }}{6 H^3} , \, \, \, \, \, {\cal Q}^{\phi}_{\, \, \,\dot\phi \phi \dot\phi}=  \frac{\dot{\phi } V_{\phi }}{4 H^5}-\frac{5 \dot{\phi }^3 V_{\phi }}{72 H^7}  ~,~\nonumber\\
& & {\cal Q}^{\dot\phi}_{\, \, \, \phi \dot\phi \dot\phi}=  \frac{5 \dot{\phi }^2 V_{\phi }^2}{72 H^7}-\frac{\dot{\phi }^2 V_{\phi \phi }}{12 H^5}-\frac{V_{\phi }^2}{12 H^5}+\frac{V_{\phi \phi }}{6 H^3}\, , \, \, \, \, \, {\cal Q}^{\phi}_{\, \, \,\dot\phi \dot\phi \dot\phi}= -\frac{5 \dot{\phi }^4}{72 H^7}+\frac{\dot{\phi }^2}{2 H^5}-\frac{1}{2 H^3} ~,~\nonumber\\
& & {\cal Q}^{\dot\phi}_{\, \, \, \phi \phi \dot\phi}= \frac{5 \dot{\phi } V_{\phi }^3}{72 H^7}-\frac{\dot{\phi } V_{\phi } V_{\phi \phi }}{4 H^5}+\frac{\dot{\phi } V_{\phi \phi \phi }}{6 H^3} \, , \, \, \, \, \,  {\cal Q}^{\dot\phi}_{\, \, \, \dot\phi \phi \dot\phi}=  \frac{5 \dot{\phi }^2 V_{\phi }^2}{72 H^7}-\frac{\dot{\phi }^2 V_{\phi \phi}}{12 H^5}-\frac{V_{\phi }^2}{12 H^5}+\frac{V_{\phi \phi }}{6 H^3} ~,~\nonumber\\
& & {\cal Q}^{\dot\phi}_{\, \, \, \dot\phi \phi \phi}=  \frac{5\dot{\phi } V_{\phi }^3}{72 H^7}-\frac{\dot{\phi } V_{\phi } V_{\phi \phi }}{4 H^5}+\frac{\dot{\phi } V_{\phi \phi \phi }}{6 H^3} \, , \, \, \, \, \, {\cal Q}^{\dot\phi}_{\, \, \, \dot\phi \dot\phi \phi}= \frac{5 \dot{\phi }^2 V_{\phi }^2}{72 H^7}-\frac{\dot{\phi }^2 V_{\phi \phi }}{12 H^5}-\frac{V_{\phi }^2}{12 H^5}+\frac{V_{\phi \phi }}{6 H^3}  ~.~\nonumber
\eea

\end{appendix}

%\clearpage
\nocite{*}
\bibliography{Tensor}
\bibliographystyle{utphys}

%%%%%%%%%%%%%%%%%%%%%%%%%%%%%%%%%%%%%%%%%%%%%%%
%%%%%%%%%%%%%%%%%%%%%%%%%%%%%%%%%%%%%%%%%%%%%%%

\end{document}